\documentclass[11pt,reqno]{amsart}
\allowdisplaybreaks[4]
\usepackage{amsthm} 

\newtheorem{example}{Example}
\usepackage{graphicx} 
\usepackage{epsfig}
\usepackage{amsthm}
\usepackage{amssymb}
\usepackage{amsmath}
\usepackage{tcolorbox}
\usepackage{cite}
\newtheorem{theorem}{Theorem}                                                                                                                           

\newtheorem{proposition}{Proposition}
\newtheorem{corollary}{Corollary}

\newtheorem{lemma}{Lemma}
\newtheorem{remark}{Remark}
\newcommand{\be}{\begin{equation}}
\newcommand{\ee}{\end{equation}}
\newcommand{\bea}{\begin{eqnarray}}
\newcommand{\eea}{\end{eqnarray}}
\newcommand{\ba}{\begin{array}}
\newcommand{\ea}{\end{array}}
\newcommand{\bean}{\begin{eqnarray*}}
\newcommand{\eean}{\end{eqnarray*}}

\newcommand{\bft}{\mathbf{t}}

\newcommand{\pa}{\partial}

\newcommand{\E}{\mathcal{E}}

\begin{document}
\title{TRANSFORMATIONS OF THE 2-COMPONENT BKP TAU FUNCTIONS}
\author{Mengyao Chen $^1$, Jipeng Cheng $^{1,2*}$, Jinbiao Wang $^1$}
\dedicatory { $^1$ School of Mathematics, China University of
Mining and Technology, Xuzhou, Jiangsu 221116, P.\ R.\ China\\$^2$ Jiangsu Center for Applied Mathematics (CUMT), Xuzhou, Jiangsu 221116, P.\ R.\ China}
\thanks{*Corresponding author. Email: chengjp@cumt.edu.cn, chengjipeng1983@163.com.}
\begin{abstract}
The 2-component BKP (2-BKP) hierarchy is an important integrable system corresponding to the infinite dimensional Lie algebras $b_{\infty}$ and $d_{\infty}$, which contains Novikov-Veselov equation and can be used to describe the total descendent potential of D type singularity. Here we firstly introduce the projections of the mixed pseudo-differential operators to rewrite the 2-BKP Lax equation in the Shiota construction, where the scalar Lax operators involving two differential operators $\pa_1$ and $\pa_2$ are used. Based upon this, the $(M_1,M_2)$-reduction of the 2-BKP hierarchy is given. After that, we give the most important result of this paper, i.e., the transformations of the 2-BKP tau functions, which are in fact the 2-BKP Darboux transformations. Here we further give the corresponding changes in the 2-BKP Lax operators. Also the corresponding results are investigated for the reduction case. Finally, the additional symmetries can be viewed as the special cases of the transformations of the 2-BKP tau functions. Besides, we discuss the Pfaffian identities of the 2-BKP tau functions by successive applications of the above transformations, which are closely related with the 2-BKP addition formulae. 
\end{abstract}
\maketitle
\tableofcontents
\section{Introduction}
The KP theory \cite{Dickey2003,Date1983,van Moerbeke1994,Ohta1988} plays an improtant role in soliton theory, which can be viewed as one universal theory, since many important nonlinear integrable equations can be found in the KP theory. Here we are more interested in the two-component BKP (2-BKP) hierarchy\cite{Shiota1989,Geng2023,Jimbo1983,van2014,Liu2011,Cheng2021,KacVandeleur1998,Kac2023}:
\begin{equation}
    \begin{aligned}
        &{\rm Res}_z z^{-1}\tau(\mathbf{t}-2[z^{-1}]_1)\tau(\mathbf{t}'+2[z^{-1}]_1)e^{\xi(t^{(1)}-t^{(1)'},z)}
\\&={\rm Res}_z z^{-1}\tau(\mathbf{t}-2[z^{-1}]_2)\tau(\mathbf{t}'+2[z^{-1}]_2)e^{\xi(t^{(2)}-t^{(2)'},z)},
\label{2bkptaubilinear}
    \end{aligned}
\end{equation}
    where ${\rm Res}_z \sum_{i}b_iz^i=b_{-1}$, $\mathbf{t}=(t^{(1)},t^{(2)})$, $t^{(a)}=(t^{(a)}_1=x_a,t^{(a)}_3,t^{(a)}_5,\ldots)$,
    $[z^{-1}]_1=([z^{-1}],0)$, $[z^{-1}]_2=(0,[z^{-1}])$, $[z^{-1}]=(z^{-1},z^{-3}/3,z^{-5}/5,\ldots)$
    and $\xi(t^{(a)},z)=\sum_{i=1}^{+\infty}t^{(a)}_{2i-1}z^{2i-1}$.
The 2-BKP hierarchy is corresponding to the infinite dimensional Lie algebra $b_{\infty}$ and $d_{\infty}$\cite{Jimbo1983,Kac2023,KacVandeleur1998,van2014}, which can be viewed as the negative extension \cite{Wang2019} of the ususal BKP hierarchy \cite{Date1983,Zabrodin2021}. 
The 2-BKP hierarchy has many important applications in mathematical physics. The famous Novikov-Veselov equation \cite{Novikov1986,Hu1996} is contained in the 2-BKP hierarchy. The total descendant potential of D type singularity \cite{Alexander2021,Cheng2021,Frenkel2010,Givental2005} is a tau function of the principle Kac-Wakimoto hierarchy\cite{Liu2011,Cheng2021,Kac1989}, which is proved to be a reduction of the 2-BKP hierarchy satisfying the string equation. In \cite{Cheng2021}, we fixed the point in the Grassmannian of the 2-BKP hierarchy that corresponds to the total descendant potential of D type singularity, which is just the Kac-Schwarz problem \cite{Kac1991} of D type, while in \cite{Alexander2021}, the results in \cite{Cheng2021} are improved, where the dilaton constraints can be removed.

Here we would like to point out that tau function and Lax equation are two important formulations for integrable hierarchies, but it is usually highly nontrivial to link these two different formulations\cite{Cheng2021-2,Cui2025,Kac1989,Kac1996}. Based upon Manakov's work in \cite{Manakov1976} and Novikov-Veselov's work on two-dimensional Schr\"{o}dinger operators \cite{Novikov1986}, Shiota expressed the 2-BKP Lax equation in \cite{Shiota1989} by two scalar Lax operators involving two differential operators $\partial_1=\partial_{x_1}$ and $\partial_2=\partial_{x_2}$, and a special operator $H=\partial_1\partial_2+\rho$ to link $\partial_1$ and $\partial_2$. Further in 
\cite{Liu2011}, the 2-BKP Lax operator can be expressed by only one differential operator $\partial_1$, where one Lax operator $L_1\in \mathcal{B}((\partial_1^{-1}))$, while another Lax operator $L_2\in \mathcal{B}((\partial_1))$ with $\mathcal{B}$ being the set of functions of $\mathbf{t}$. In \cite{Geng2023}, the 2-BKP hierarchy is described by the dressing operators involving $\partial_1$ and $\partial_2$. In \cite{van2014}, the $2\times 2$ matrix pseudo-differential operators are used to describe the Lax operators for the system involving both the 2-BKP hierarchy and the 2-component modified BKP (2-mBKP) hierarchy, and the string equation and $W$ constraints are also discussed. In this paper, we will discuss the transformations of the 2-BKP tau functions and give the corresponding changes in the setting of the 2-BKP Lax structure. 

\subsection{The Lax structure of the 2-BKP hierarchy and its reduction}

Firstly, we rewrite the 2-BKP Lax equation involving $\partial_1$ and $\partial_2$ in the Shiota construction \cite{Shiota1989} by introducing the projections in Subsection \ref{sec:Lax Representation},
\begin{align*}
  \pi_a:\mathcal{E}_{(a)}=\mathcal{E}_{(a)}^0\oplus\mathcal{E}_{(a)}H\rightarrow\mathcal{E}_{(a)}^0, \quad a=1,2,
\end{align*}
where $\mathcal{E}_{(a)}:=\mathcal{B}[\partial_{3-a}]((\partial^{-1}_a)),$
$\mathcal{E}_{(a)}^0:=\mathcal{B}((\partial^{-1}_a))$. The 2-BKP Lax equation can be expressed by a triple $(L_1,L_2,H)$, with the Lax operators $L_a:=\partial_a+\sum_{i=1}^{+\infty}u_{a,i}\partial_a^{-i}$ and the special operator $H:=\partial_1\partial_2+\rho$ satisfying
\begin{equation}\label{2bkplaxtriple}
\begin{aligned}
  &L^*_a=-\partial_aL_a\partial_a^{-1},\\
  &\partial_{a,n}L_b=[\pi_{b}(B_n^{(a)}),L_{b}],\\
  &\partial_{a,n}H=-(B^{(a)}_{n})^*H-HB^{(a)}_{n},\quad a,b=1,2,
\end{aligned}
\end{equation}
where $B^{(a)}_{n}:=(L^n_a)_{a,\geq 0}$, $(\sum_{i}b_i\partial_a^i)^*:=\sum_{i}(-\partial_a)^i b_i$, $(\sum_{i}b_i\partial_a^i)_{a,\geq 0}:=\sum_{i\geq 0}b_i\partial_a^i$ and $\partial_{a,n}=\partial_{t^{(a)}_n}$. In this paper, $n$ is always the odd positive integer.
Here the 2-BKP Lax operators $L_1$ and $L_2$ are related with the 2-BKP tau function $\tau(\mathbf{t})$ by the way below,
\begin{equation}\label{laxexpression}
  \begin{aligned}
  &L_a(\mathbf{t},\partial_{a})=\sum_{j,l=0}^{+\infty}(-1)^{l}
  \frac{h_j(-2\widetilde{\partial}_{t^{(a)}})\tau(\mathbf{t})}{\tau(\mathbf{t})}\partial_{a}^{-j-l}
  \frac{h_j(-2\widetilde{\partial}_{t^{(a)}})\tau(\mathbf{t})}{\tau(\mathbf{t})}\partial_{a}, \quad a=1,2,
\end{aligned}
\end{equation}
where $\widetilde{\partial}_{t^{(a)}}=(\partial_{t_1^{(a)}},\frac{1}{3}\partial_{t_3^{(a)}},\frac{1}{5}\partial_{t_5^{(a)}},\ldots)$
and $h_j(\boldsymbol{y})$ is determined by $e^{\xi(\boldsymbol{y},\lambda)}=\sum_{j=0}^{+\infty}h_j(\boldsymbol{y})\lambda^{j}$ with $\boldsymbol{y}=(y_1,y_3,\ldots)$. 
Also $\rho$ in $H=\partial_1\partial_2+\rho$ is given by 
\begin{align}\label{rho}
\rho(\mathbf{t})=2\partial_1\partial_2\big(\log\tau(\mathbf{t})\big).
\end{align}

In Subsection \ref{bilineartolax} and Subsection \ref{laxtobilinear}, we will show the equivalence of the 2-BKP Lax equation (\ref{2bkplaxtriple}) and the 2-BKP bilinear equation (\ref{2bkptaubilinear}), where we should remark that the equivalence of the 2-BKP bilinear equation and the dressing structure is contained in \cite{Liu2011,Geng2023}, while it is new for the equivalence of the 2-BKP dressing structure and Lax equations.

After the preparation above, we discuss the reduction of the 2-BKP hierarchy in Subsection \ref{reductions2bkp}. In fact, the $(M_1,M_2)$-reduction of the 2-BKP hierarchy can be given by the following constraints on the 2-BKP Lax operators $L_1$ and $L_2$:
\begin{align}\label{constraintm1m2}
  (L_a^{M_a})_{a,\leq 0}=\pi_a(B_{M_{3-a}}^{(3-a)}),\quad a=1,2,
\end{align}
where $B_{M_{3-a}}^{(3-a)}=(L_{3-a}^{M_{3-a}})_{3-a,\geq0}$,  $\big(\sum_{i}b_i\pa_a^i\big)_{a,[0]}:=b_0$, $\big(\sum_{i}b_i\pa_a^i\big)_{a,\leq0}:=\sum_{i\leq0}b_i\pa_a^i$. Here $M_1,M_2\in\mathbb{Z}_{>0}$ and $M_1+M_2$ is an even integer. 
Since $(L_1^n)_{1,[0]}=(L_2^n)_{2,[0]}=0$ for $n$ odd, $M_1+M_2$ can not be odd (please see \cite{Date1983,Zabrodin2021}). In terms of tau functions, the $(M_1,M_2)$-reduction of the 2-BKP hierarchy is given by 
\begin{align}\label{bilinearm1m2reduction}
  {\rm Res}_z \sum_{a=1}^{2}(-1)^az^{M_a-1}\tau(\mathbf{t}-2[z^{-1}]_a)\tau(\mathbf{t}'+2[z^{-1}]_a)e^{\xi(t^{(a)}-t^{(a)'},z)}=0.
\end{align}
In fact, it can be proved that (\ref{constraintm1m2}) is equivalent to (\ref{bilinearm1m2reduction}). 
From (\ref{constraintm1m2}), we can know the $(M_1,M_2)$-reduction of the 2-BKP hierarchy can be expressed by a Lax operator pair $(\mathcal{L},H)$ with the Lax operator,
   \begin{align}\label{mathcall}
    \mathcal{L}=
\begin{split}
\left \{
\begin{array}{ll}
      \sum_{a=1}^{2}\Big(\pa_a^{M_a}+\sum_{l=1}^{\frac{M_a-3}{2}}(v_{a,l}\pa_a^{2l}
    +\pa_a^{2l}v_{a,l})\pa_a+v_{a,0}\pa_a\Big),\quad M_{a}~\rm{is~odd},\\
     \sum_{a=1}^{2}\Big(\pa_a^{M_a}+\sum_{l=1}^{\frac{M_a-2}{2}}(\widetilde{v}_{a,l}\pa_a^{2l-1}
    +\pa_a^{2l-1}\widetilde{v}_{a,l})\pa_a\Big)+\widetilde{v}_{0},\quad M_{a}~\rm{is~even},
\end{array}
\right.
\end{split}
\end{align}
where $\mathcal{L}$ and $H$ satisfy 
\begin{align*}
  &H\mathcal{L}=(-1)^{M_a}\mathcal{L}^*H,\quad \partial_{a,n}H=-(B^{(a)}_{n})^*H-HB^{(a)}_{n}, \\
  &\pa_{a,n}\mathcal{L}=[\pi_1\big(B_n^{(a)}\big),\pi_1(\mathcal{L})]_{1,\geq1}+[\pi_2\big(B_n^{(a)}\big),\pi_2(\mathcal{L})]_{2,\geq0}, \quad a=1,2.
\end{align*}
Here $B_n^{(a)}=\big(\pi_a(\mathcal{L})^{\frac{n}{M_a}}\big)_{a,\geq0}$. Notice that by the similar method in Lemma \ref{welldefined} (in Subsection \ref{laxtobilinear}), we can find 
\[\pi_a(\mathcal{L})^*=-\pa_a\cdot\pi_a(\mathcal{L})\cdot\pa_a^{-1}.\]
\begin{remark}
 When $M_1$ and $M_2$ are even positive integers, the above $(M_1,M_2)$-reduction of the 2-BKP hierarchy is just the $\big(\emptyset,(M_1,M_2)\big)$-reduced DKP hierarchy in \cite{Kac2023,KacVandeleur1998} corresponding to the affine Lie algebra $\widehat{\mathfrak{so}}_{M_1+M_2}$ with the conjugacy class of the pair of partition $\emptyset$ and $(M_1,M_2)$. When $M_1$ and $M_2$ are odd positive integers, the $(M_1,M_2)$-reduction means the system of 2-BKP hierarchy does not depend on $t_{M_1k}^{(1)}+t_{M_2k}^{(2)}$ $(k\geq1, odd)$ (please see Remark \ref{oddm1m2} in Subsection \ref{reductions2bkp} for more details).
\end{remark}

\subsection{The transformations of 2-BKP tau functions}
In Section \ref{2bkpTransformations}, we will consider the transformations of the 2-BKP tau functions. For the 2-BKP eigenfunction $q$ satisfying 
\begin{align}\label{eigenfunction2bkp}
 \partial_{a,n}q=B^{(a)}_{n}(q),\quad H(q)=0,
\end{align}
 we have the following theorem.
\begin{theorem}\label{mainqtau}
  Given the tau function $\tau(\mathbf{t})$ and the eigenfunction $q(\mathbf{t})$ of the 2-BKP hierarchy, $$\tau^{[1]}(\mathbf{t}):=q(\mathbf{t})\tau(\mathbf{t})$$ is another 2-BKP tau function, i.e., satisfying the 2-BKP bilinear equation (\ref{2bkptaubilinear}).

  Further if $(L_1,L_2,H)$ is the 2-BKP Lax triple corresponding to $\tau(\mathbf{t})$ by (\ref{laxexpression}) and (\ref{rho}), then the new tau function $\tau^{[1]}(\mathbf{t})$ will be corresponding to the new 2-BKP Lax triple $(L^{[1]}_1,L^{[1]}_2,H^{[1]})$ given by
   \begin{align*}
     &L^{[1]}_a=T_a[q]L_aT_a[q]^{-1},\quad H^{[1]} =\partial_1T_1[q]\partial_1^{-1}HT_1[q]^{-1}
     =\partial_2T_2[q]\partial_2^{-1}HT_2[q]^{-1},
   \end{align*}
   where the operator $T_a[q]$ is given by
   \begin{align*}
     T_a[q]=1-2q^{-1}\partial_a^{-1}q_{x_a},\quad a=1,2.
   \end{align*}
\end{theorem}

\begin{remark}
The transformation in Theorem \ref{mainqtau} is quite basic, which is in fact the Darboux transformation of the 2-BKP hierarchy. Thus, $T_a[q]$ is the 2-BKP Darboux transformation operator, which shares the same form with the usual BKP case\cite{Cheng2014,He2007,Nimmo1995}. There are many important applications for this kind of transformations. For example, the additional symmetries\cite{Dickey2003,van Moerbeke1994}, which plays an important role in the so-called string equation and the generalized Virasoro constraints in the matix models of the 2-d quantum gravity (see \cite{Dickey2003,van Moerbeke1994} and their references).
\end{remark}
\begin{corollary}\label{qtauistau}
Under the same conditions as Theorem \ref{mainqtau}, if $\widetilde{q}(\mathbf{t})$ is another 2-BKP eigenfunction corresponding to $\tau(\mathbf{t})$, then
\begin{align*}
 \widetilde{q}^{[1]}=q^{-1}\Omega(\widetilde{q},q)
\end{align*}
is the eigenfunction corresponding to $\tau^{[1]}(\mathbf{t})$ by (\ref{laxexpression})(\ref{rho})(\ref{eigenfunction2bkp}), where $\Omega(\widetilde{q},q)$ is the squared eigenfunction potential (SEP) defined by 
$$\Omega(\widetilde{q},q):=\partial_1^{-1}(q\widetilde{q}_{x_1}-\widetilde{q}q_{x_1})
=-\partial_2^{-1}(q\widetilde{q}_{x_2}-\widetilde{q}q_{x_2}).$$
Meanwhile, $\Omega(\widetilde{q},q)\tau$ is also another 2-BKP tau function, satisfying the 2-BKP bilinear equation (\ref{2bkptaubilinear}). 
 \end{corollary}
 \begin{remark}
  Due to $H(q)=0$, we can find $(q_2q_{1x_1}-q_1q_{2x_1})_{x_2}=-(q_2q_{1x_2}-q_1q_{2x_2})_{x_1}$, thus the definition of $\Omega(q_1,q_2)$ is reasonable (please see Lemma \ref{omegadef1} in Subsection \ref{sec:Lax Representation} for more details).
 \end{remark}
\begin{proposition}\label{m1m2redction}
 For the $(M_1,M_2)$-reduction of the 2-BKP hierarchy, assume the Lax operator $\mathcal{L}$ of the form (\ref{mathcall}), along with the corresponding eigenfunction $q(\bft)$ and the tau function $\tau(\bft)$. If we have $$\mathcal{L}\big(q(\bft)\big)=c\cdot q(\bft)$$ 
 for some constant $c$, then $\tau^{[1]}(\bft)=q(\bft)\tau(\bft)$ will be the new tau function of the $(M_1,M_2)$-reduction of the 2-BKP hierarchy, i.e.,
 \begin{align*}
  {\rm Res}_z \sum_{a=1}^{2}(-1)^az^{M_a-1}q(\mathbf{t}-2[z^{-1}]_a)\tau(\mathbf{t}-2[z^{-1}]_a)q(\mathbf{t'}+2[z^{-1}]_a)\tau(\mathbf{t}'+2[z^{-1}]_a)e^{\xi(t^{(a)}-t^{(a)'},z)}=0.
 \end{align*}
Under the transformation $\tau(\bft)\rightarrow\tau^{[1]}(\bft)=q(\bft)\tau(\bft)$, the Lax operator $\mathcal{L}$ will become
\begin{align*}
  \mathcal{L}^{[1]}=\big(T_1[q]\pi_1(\mathcal{L})T_1[q]^{-1}\big)_{1,\geq1}+\big(T_2[q]\pi_2(\mathcal{L})T_2[q]^{-1}\big)_{2,\geq0},
\end{align*}
and $H$ becomes $H^{[1]} =\partial_1T_1\partial_1^{-1}HT_1^{-1}=\partial_2T_2\partial_2^{-1}HT_2^{-1}
$.
\end{proposition}
 \begin{corollary}\label{mnreduction1} 
   Under the same condition of Proposition \ref{m1m2redction}, assume $\widetilde{q}$ to be another 2-BKP eigenfunction, if $\mathcal{L}(q)=c\cdot q$ and $\mathcal{L}(\widetilde{q})=\widetilde{c}\cdot \widetilde{q}$ for two constants $c$ and $\widetilde{c}$, then $\Omega(\widetilde{q},q)\tau$ is the new tau function of the $(M_1,M_2)$-reduction of the 2-BKP hierarchy.
 \end{corollary}

\subsection{Applications in the additional symmetries and Pfaffian identities}\label{introapplications}
Here we would like to consider some applications of the transformations of the 2-BKP tau functions (see Theorem \ref{mainqtau}).

Firstly, the additional symmetries of the 2-BKP hierarchy can be viewed as the special cases of $\tau\rightarrow \Omega(\widetilde{q},q)\tau$. In fact, if we introduce the following vertex operators\cite{Date1983,Tu2007,Wu2013}:
\begin{align*}
    &X_{ab}(\lambda,\mu):=\varepsilon_{ab}(\lambda,\mu)e^{\xi(t^{(a)},\mu)-\xi(t^{(b)},\lambda)}e^{-2\xi(\widetilde{\partial}_{t^{(a)}},\mu^{-1})+2\xi(\widetilde{\partial}_{t^{(b)}},\lambda^{-1})},\quad a,b=1,2,
 \end{align*}
 where $\varepsilon_{ab}(\lambda,\mu)=\frac{\mu-\lambda}{\mu+\lambda}$ if $a=b$, and $\varepsilon_{ab}(\lambda,\mu)=1$ if $a\neq b$.
And the 2-BKP wave function:
\[
\psi_a(\mathbf{t},z):=\frac{\tau(\mathbf{t}-2[z^{-1}]_a)}{\tau(\mathbf{t})}e^{\xi(t^{(a)},z)},\quad a=1,2,
\] then we have the following lemma.
\begin{lemma}\label{asvmformual}
  Given the 2-BKP tau function $\tau(\mathbf{t})$,
  \begin{align*}
    &\frac{X_{ab}(\lambda,\mu)\tau(\mathbf{t})}{\tau(\mathbf{t})}=(-1)^a\Omega\big(\psi_b({\mathbf{t}},-\lambda),\psi_a({\mathbf{t}},\mu)\big),\quad a,b=1,2.
  \end{align*}
\end{lemma}

The following additional flow $\partial^*_{ab}$ is defined by the action on 2-BKP wave function $\psi_c(\mathbf{t},z)$
 \begin{align*}
   &\partial^*_{ab}\psi_{c}(\mathbf{t},z):=2
    \Big(\psi_a(\mathbf{t},\mu)\partial_{c}^{-1}\psi_{b,x_c}(\mathbf{t},-\lambda)
    -\psi_b(\mathbf{t},-\lambda)\partial_{c}^{-1}\psi_{a,x_c}(\mathbf{t},\mu)
    \Big)\big(\psi_{c}(\mathbf{t},z)\big),\quad a,b,c=1,2,
  \end{align*}
then we have the following theorem.
\begin{theorem}\label{widetildetau}
   Given the 2-BKP tau function $\tau(\mathbf{t})$,  $$\widetilde{\tau}_{ab}:=\tau+C_{ab}X_{ab}(\lambda,\mu)\tau,\quad a,b=1,2$$ is also the 2-BKP tau function, where $C_{ab}$ is some constant. And in particular,
   \begin{align}\label{paabpaic}
 &\partial^*_{ab}\psi_{c}(\bft,z)
 =\psi_c(\bft,z)(e^{-2\xi(\widetilde{\partial}_{t^{(c)}},z)}-1)
 \frac{X_{ab}(\lambda,\mu)\tau(\bft)}{\tau(\bft)}, \quad a,b,c=1,2.
  \end{align}
\end{theorem}
\begin{remark}
  Here $\pa^*_{ab}$ can be viewed as the generator of the additional symmetries for the 2-BKP hierarchy, where the additional symmetry is a kind of symmetry explicitly depending on time flows. And the formula (\ref{paabpaic}) is the so-called Adler-Shiota-van Moerbeke formula \cite{Adler1995,Dickey1995,Tu2007} which connects the actions of additional symmetries on the wave functions and tau functions. In particular when $a=b=c$, the corresponding results have been obtained in \cite{Wu2013}, but other cases are not considered.
\end{remark}
\begin{proposition}\label{mnreductionvetex}
  Under the same conditions of Theorem \ref{widetildetau}, when
  \begin{align*}
     \mu^{M_a}=\lambda^{M_b},
  \end{align*}
$\widetilde{\tau}_{ab}=\big(1+C_{ab}X_{ab}(\lambda,\mu)\big)\tau$ is the tau function of the $(M_1,M_2)$-reduction of the 2-BKP hierarchy. 
\end{proposition}
Another application of Theorem \ref{mainqtau} is the Pfaffian identities of the 2-BKP tau function, which is derived by using the BKP Darboux operator $T[q]=1-2q^{-1}\partial^{-1}q_{x}$
\cite{Cheng2014,He2007,Nimmo1995}.

\begin{theorem}\label{pfaffiantau}
 Given the 2-BKP tau function $\tau(\mathbf{t})$,
\begin{align*}
\begin{split}
&\frac{\tau(\mathbf{t}-2\sum_{q=1}^{N_1}[\lambda_q^{-1}]_1-2\sum_{q=1}^{N_2}
    [\mu_q^{-1}]_2)}{\tau(\mathbf{t})} \prod_{\substack{1\leq i<j\leq N_1\\ 1\leq s<l\leq N_2}}\frac{\lambda_i-\lambda_j}{\lambda_i+\lambda_j}
\frac{\mu_l-\mu_s}{\mu_l+\mu_s}
   \\&\qquad=\left \{
\begin{array}{ll}
    \mathrm{Pf}
\begin{pmatrix}
    E & F \\
    -F^{T} & G
\end{pmatrix}_{m\times m},                    &m~\rm{is~even},\\
    \mathrm{Pf}
\begin{pmatrix}
    E & F & C \\
    -F^{T} &G &D \\
   -C^{T} & -D^{T} & 0\\
\end{pmatrix}_{(m+1)\times (m+1)},                                 & m~\rm{is~odd}, 
\end{array}
\right.
\end{split}
\end{align*}
where $\mathrm{Pf}$ is the Pfaffian \big(see (\ref{pfdef}) in Appendix \ref{app:first}\big), $|\lambda_{N_1}|<|\lambda_{N_1-1}|< \ldots <|\lambda_1|$ and $|\mu_{N_2}|<|\mu_{N_2-1}|< \ldots <|\mu_1|$. Here the matrices $E, F, G$ and $C, D$ are given by
    \begin{align*}
        &E_{N_1\times N_1}=\bigg(\frac{\lambda_{i}-\lambda_{j}}{\lambda_{i}+\lambda_{j}}\frac{\tau(\mathbf{t}-2[\lambda_{i}^{-1}]_1-2[\lambda_{j}^{-1}]_1)}{\tau(\mathbf{t})}\bigg)
       _{1\leq i,j\leq N_1},\\
       &G_{N_2\times N_2}=\bigg(\frac{\mu_{j}-\mu_{i}}{\mu_{j}+
       \mu_{i}}\frac{\tau(\mathbf{t}-2[\mu_{i}^{-1}]_2
       -2[\mu_{j}^{-1}]_2)}{\tau(\mathbf{t})}\bigg)
       _{1\leq i,j\leq N_2},\\
       &F_{N_1\times N_2}=\bigg(\frac{\tau(\mathbf{t}-2[\lambda_{i}^{-1}]_1-
       2[\mu_{j}^{-1}]_2)}{\tau(\mathbf{t})}\bigg)
       _{1\leq i\leq N_1, 1\leq j\leq N_2},\\     
        &C_{N_1\times 1}=\bigg(\frac{\tau(\mathbf{t}-2[\lambda_{i}^{-1}]_1)}{\tau(\mathbf{t})}\bigg)
       _{1\leq i\leq N_1},
     \quad D_{N_2\times 1 }=\bigg(\frac{\tau(\mathbf{t}-2[\mu_{i}^{-1}]_2)}{\tau(\mathbf{t})}\bigg)_
     {1\leq i\leq N_2}. 
    \end{align*}
\end{theorem}
\begin{remark}
  When $N_1=0$ or $N_2=0$, we can obtain the Pfaffian identities for the usual BKP hierarchy\cite{Shigyo2013,Zabrodin2025}, which is related with the BKP addition formulae \cite{Shigyo2013} by considering $Pl\ddot{u}cker$ relation of the Pfaffians. Thus we believe the Pfaffian identities in Theorem \ref{pfaffiantau} should also related with the addition formulae of the 2-BKP hierarchy in \cite{Gao2016}.
\end{remark}
 
\section{The Lax equation of the 2-BKP hierarchy}
In this section, we firstly introduce the projection $\pi_a:\mathcal{E}_{(a)}\rightarrow\mathcal{E}_{(a)}^0$ ($a=1,2$) and then
rewrite the 2-BKP Lax equation in the Shiota construction. Finally, the $(M_1,M_2)$-reduction of the 2-BKP hierarchy is given.
\subsection{Projections of the mixed pseudo-differential operators}\label{sec:Lax Representation}
Firstly, let us recall the following definitions:
\begin{align*}
\mathcal{E}_{(a)}:=\mathcal{B}[\partial_{3-a}]((\partial^{-1}_a)),\quad
\mathcal{E}_{(a)}^0:=\mathcal{B}((\partial^{-1}_a)),\quad a=1,2,
\end{align*}
where $\mathcal{B}$ is the set of functions of $\mathbf{t}$, and the element of $\E_{(a)}$ is called the mixed pseudo-differential operator. For the special operator $H=\partial_1\partial_2+\rho$, we have the following proposition.
\begin{proposition}\label{proH1}
$\mathcal{E}_{(a)}$ has the following direct sum decomposition:
  \begin{align*}
\mathcal{E}_{(a)}=\mathcal{E}^0_{(a)}\oplus\mathcal{E}_{(a)}H,\quad a=1,2.
\end{align*}
\begin{proof}
Here, we only prove the case of $a=1$, since $a=2$ is quite similar to $a=1$. Firstly, let us show $\mathcal{E}^0_{(1)}\cap\mathcal{E}_{(1)}H=\{0\}$. Notice that $h\in\mathcal{E}^0_{(1)}\cap\mathcal{E}_{(1)}H$ can be written into the form below:
\begin{align*}
 h&=\sum_{i\leq M}\sum_{0\leq j\leq N}b_{i,j}\partial_1^{i}\partial_2^{j}H.
\end{align*}
By inserting $H=\partial_1\partial_2+\rho$ into $h$, we can obtain:
\begin{align*}
 h&=\sum_{i\leq M+1}\sum_{j\leq N+1}b_{i-1,j-1}\partial_1^{i}\partial_2^{j}+\sum_{i\leq M}\sum_{j\leq N}\sum_{s=i}^{M}\sum_{k=j}^{N}b_{s,k}C_{k}^{k-j}C_{s}^{s-i}\cdot\partial_1^{k-j}\partial_2^{s-i}(\rho)\cdot\partial_1^{i}\partial_2^{j},
\end{align*}
where we assume $b_{i,j}=0$ for $j<0$ and $C_{k}^{l}=\frac{k(k-1)\cdots(k-l+1)}{l!}$. Because $h\in\mathcal{E}_{(1)}^0$, it implies that the coefficients of $\partial_2^{j}
$ $(j\leqslant N+1)$ vanish and thus
\begin{align}
&b_{i-1,N}=0,\quad i\leq M+1,\label{bij1}\\
& b_{M,j-1}=0,\quad j\leq N,\label{bij2}\\
&b_{i-1,j-1}+\sum_{s=i}^{M}\sum_{k=j}^{N}b_{s,k}C_{k}^{k-j}C_{s}^{s-i}\cdot\partial_1^{k-j}\partial_2^{s-i}(\rho)=0, \quad j\leq N,\quad i\leq M. \label{bij3} 
\end{align}
Notice that if set $j=N$ in (\ref{bij3}), then $$b_{i-1,N-1}+\sum_{s=i}^{M}b_{s,N}C_{s}^{s-i}\cdot\partial_2^{s-i}(\rho)=0,$$
So by (\ref{bij1}), we can derive $b_{i-1,N-1}=0$ for $i\leq M$. Further by (\ref{bij2}), we can obtain $b_{i,N-1}=0$ for $i\leq M$. Continue the above procedure, we finally obtain $b_{i,j}=0$ for $j\leq N$ and $i\leq M$, which means $h=0$.

Next let us prove $\mathcal{E}_{(1)}=\mathcal{E}^0_{(1)}+\mathcal{E}_{(1)}H$. In fact, we only need to show 
\begin{align}\label{pidecomposition1}
  \partial_2^k\partial_1^l\in \mathcal{E}^0_{(1)}+\mathcal{E}_{(1)}H,\quad k\geq0,\quad l\in \mathbb{Z}.
\end{align}
If we assume the statement (\ref{pidecomposition1}) holds for $k\leq q$, then there exist $A_{q,l}\in\mathcal{E}^0_{(1)}$, $B_{q,l}\in\mathcal{E}_{(1)}$ such that
\begin{align*}
   \partial_2^q\partial_1^l=A_{q,l}+B_{q,l}H.
 \end{align*}
When $k=q+1$, we can find
\begin{align*}
  \partial_2^{q+1}\partial_1^l&=\partial_2\cdot(A_{q,l}+B_{q,l}H)
  \\&=\partial_2(A_{q,l})-A_{q,l}\partial_2^{-1}\rho+(A\pa_1^{-1}+\partial_2B_{q,l})H.
\end{align*}
So (\ref{pidecomposition1}) is correct for $k=q+1$.
\end{proof}
\end{proposition}
\begin{remark}
  The result of Proposition \ref{proH1} is given in \cite{Shiota1989}, but there is no proof.
\end{remark}

Thus we can define the projection from $\mathcal{E}_{(a)}$ to $\mathcal{E}_{(a)}^0$, that is
\begin{align*}
  \pi_a:\mathcal{E}_{(a)}=\mathcal{E}_{(a)}^0\oplus\mathcal{E}_{(a)}H\rightarrow\mathcal{E}_{(a)}^0.
\end{align*}
By the definition of $\pi_{a}$, we have
\begin{align*}
 \pi_a(\partial_{3-a})=-\partial_a^{-1}\rho.
\end{align*}
Further we have the following important results for $\pi_a$.
\begin{lemma}\label{piaaction}
$\pi_{a}$ satisfies the following relations for $k\geq1$, $l\in\mathbb{Z}$, $a=1,2$,
\begin{align*}
&\pi_{a}(\partial_{3-a}^{k}\partial_a^{l})=\partial_{3-a}(\pi_{a}(\partial_{3-a}^{k-1}\partial_a^{l}))+
\pi_{a}(\partial_{3-a}^{k-1}\partial_a^{l})\pi_a(\partial_{3-a})=\partial_{a}\cdot\pi_{a}(\partial_{3-a}^{k}\partial_a^{l-1}).
 \end{align*}
\begin{proof}
  Here we only give the proof for $a=2$.
Based on $\mathcal{E}_{(2)}=\mathcal{E}^0_{(2)}\oplus\mathcal{E}_{(2)}H$, there exist $A,B\in\mathcal{E}_{(2)}$ such that $\partial_1^k\partial_2^{l-1}=\pi_2(\partial_1^k\partial_2^{l-1})+AH$ and $\partial_1^{k-1}\partial_2^{l}=\pi_2(\partial_1^{k-1}\partial_2^{l})+BH$. Then we have
\begin{align*}
\pi_{2}(\partial_{1}^{k}\partial_2^{l})
&=\pi_{2}(\partial_{2}\partial_{1}^{k}\partial_2^{l-1})
=\pi_{2}\big(\partial_{2}\cdot\pi_{2}(\partial_{1}^{k}\partial_2^{l-1})+\partial_{2}\cdot AH\big)=\partial_{2}\cdot\pi_{2}(\partial_{1}^{k}\partial_2^{l-1}),
\end{align*}
and
\begin{align*}
\pi_{2}(\partial_{1}^{k}\partial_2^{l})
&=\pi_{2}(\partial_{1}\partial_{1}^{k-1}\partial_2^{l})
=\pi_{2}\big(\partial_{1}\cdot\pi_{2}(\partial_{1}^{k-1}\partial_2^{l})+BH\big)\\
&=\pi_{2}\big(\partial_{1}(\pi_{2}(\partial_{1}^{k-1}\partial_2^{l}))+\pi_2(\partial_{1}^{k-1}\partial_2^{l})\cdot\partial_1\big)\\
&=\partial_{1}\big(\pi_{2}(\partial_{1}^{k-1}\partial_2^{l})\big)+\pi_{2}(\partial_{1}^{k-1}\partial_2^{l})\pi_{2}(\partial_{1}).
\end{align*}
\end{proof}
\end{lemma}
\begin{proposition}\label{2-bkplaxPi}
The projection $\pi_{a}$ can be computed by
  \begin{align}\label{2-bkplaxPieq}
\pi_{a}(\partial_{3-a}^k)=(\partial_{3-a}^k\iota_{\partial_{3-a}^{-1}}H^{-1}\partial_{3-a})_{3-a,[0]}\partial_a,\quad a=1,2,\quad k>0.
\end{align}
Here we have assumed $\iota_{\partial_2^{-1}}H^{-1}\partial_2=\partial_1^{-1}+\sum_{j=1}^{+\infty}\partial_2^{-j}\cdot V_j(\partial_1)$, $\iota_{\partial_1^{-1}}H^{-1}\partial_1=\partial_2^{-1}+\sum_{j=1}^{+\infty}\partial_1^{-j}\cdot R_j(\partial_2)$.
\begin{proof}
Since the cases of $a=1$ and $a=2$ are the same, here we only prove (\ref{2-bkplaxPieq}) for $a=1$, that is,
\begin{align*}
  \pi_{1}(\partial_2^k)=(\partial_2^k\iota_{\partial_2^{-1}}H^{-1}\partial_2)_{2,[0]}\partial_1.
\end{align*}
Firstly notice that $\iota_{\partial_2^{-1}}H^{-1}\partial_2=\partial_1^{-1}+\sum_{j=1}^{+\infty}\partial_2^{-j}V_j$, which means:
\begin{align}\label{partial2h}
 \partial_2=(\partial_1\partial_2+\rho)\big(\partial_1^{-1}+\sum_{j=1}^{+\infty}\partial_2^{-j}V_j\big).
\end{align} 
By comparing the coefficients of $\partial_2^j$ in (\ref{partial2h}), we can obtain
\begin{align}
&V_1=-\partial_1^{-1}\rho\partial_1^{-1},  \quad V_{j+1}=-\sum_{l=0}^{j-1}C_{j-1}^l\partial_1^{-1}\partial_2^l(\rho)V_{j-l},\quad j\geq1.\label{Vjprese1}
\end{align}
Also by considering the negative-order terms of $\partial_2$ in (\ref{partial2h}), we have
\begin{align}
&\sum_{j=2}^{+\infty}\partial_2^{1-j}V_j=
-\partial_1^{-1}\rho\sum_{j=1}^{+\infty}\partial_2^{-j}V_j.\label{Vjprese2}
\end{align}

 If we set
 $A_k=(\partial_{2}^k\iota_{\partial_2^{-1}}H^{-1}\partial_{2})_{2,[0]}\partial_1,$ then 
 \begin{align}\label{akexpress}
  A_k=\sum_{j=1}^{k}\partial_2^{k-j}(V_j)\partial_1.
 \end{align}
 Now let us prove $\pi_{1}(\partial_2^k)=A_k$ by the induction on $k$. It is obviously correct by $\pi_1(\partial_2)=-\partial_1^{-1}\rho$ when $k=1$. If assume $\pi_{1}(\partial_2^k)=A_k$ holds, then let us try to prove  
 $\pi_{1}(\partial_2^{k+1})=A_{k+1}.$ 
 Based on Lemma \ref{piaaction}, we only need to show
 \begin{align}\label{akdefinition}
A_{k+1}=\partial_2(A_k)-A_k\partial_1^{-1}\rho.
 \end{align}
 Further substituting (\ref{akexpress}) into (\ref{akdefinition}), we have $V_{k+1}\partial_1=-\sum_{j=1}^{k}\partial_2^{k-j}(V_j)\rho$. Then by (\ref{Vjprese1}), we now only need to prove
\begin{align}\label{proofmaental1}
\sum_{l=0}^{k-1}C_{k-1}^{l}\partial_1^{-1}\partial_2^{l}(\rho)V_{k-l}\partial_1=\sum_{j=1}^k\partial_2^{k-j}(V_j)\rho.
\end{align}
 And we can find (\ref{proofmaental1}) is equivalent to
 \begin{align}
  &{\rm Res}_{\partial_2}(\partial_2^{k-1}\partial_1^{-1}\rho \iota_{\partial_2^{-1}}H^{-1}\partial_2\partial_1)={\rm Res}_{\partial_2}(\partial_2^k\iota_{\partial_2^{-1}}H^{-1}\rho), \quad k\geq1.\label{proofmaental2}
  \end{align}
In fact, 
 \begin{align*}
\mathrm{LHS}~\mathrm{of}~(\ref{proofmaental2})=&{\rm Res}_{\partial_2}\big(\partial_2^{k-1}\partial_1^{-1}\rho (\partial_1^{-1}+\sum_{j=1}^{+\infty}\partial_2^{-j}V_j)\partial_1\big)={\rm Res}_{\partial_2}\partial_2^{k-1}\partial_1^{-1}\rho \sum_{j=1}^{+\infty}\partial_2^{-j}V_j\partial_1\\
=&{\rm Res}_{\partial_2}\big(\sum_{l=0}^{k-1}C_{k-1}^{l}\partial_1^{-1}\partial_2^{l}(\rho)\partial_2^{k-l-1}\sum_{j=1}^{+\infty}\partial_2^{-j}V_j\partial_1\big)
=\sum_{l=0}^{k-1}C_{k-1}^{l}\partial_1^{-1}\partial_2^{l}(\rho)V_{k-l}\partial_1,\\
\mathrm{RHS}~\mathrm{of}~(\ref{proofmaental2})=&{\rm Res}_{\partial_2}\partial_2^k(\partial_1^{-1}+\sum_{j=1}^{+\infty}\partial_2^{-j}V_j)\partial_2^{-1}\rho={\rm Res}_{\partial_2}\partial_2^k\sum_{j=1}^{+\infty}\partial_2^{-j}V_j\partial_2^{-1}\rho
=\sum_{j=1}^k\partial_2^{k-j}(V_j)\rho.
\end{align*}
Thus $(\ref{proofmaental1})\Leftrightarrow(\ref{proofmaental2})$. Further, (\ref{proofmaental2}) is equivalent to
\begin{align}\label{proofmaental3}
&(\partial_1^{-1}\rho \cdot\iota_{\partial_2^{-1}}H^{-1}\partial_2\partial_1)_{2,<0}
=(\partial_2\cdot\iota_{\partial_2^{-1}}H^{-1}\rho)_{2,<0}.
\end{align}

 Next inserting $\iota_{\partial_2^{-1}}H^{-1}\partial_2=\partial_1^{-1}+\sum_{j=1}^{+\infty}\partial_2^{-j}V_j$ into (\ref{proofmaental3}) and using (\ref{Vjprese2}), we have
 \begin{align*}
   \partial_1^{-1}\rho\cdot\sum_{j=1}^{+\infty}\partial_2^{-j}V_j\cdot\partial_1=\sum_{j=1}^{\infty}\partial_2^{1-j}V_j\partial_2^{-1}
\rho=V_1\partial_2^{-1}\rho-\partial_1^{-1}\rho\sum_{j=1}^{+\infty}\partial_2^{-j}V_j.
 \end{align*}
  After multiplying both sides by $\rho^{-1}\partial_1$, we have
 \begin{align*}
 \sum_{j=1}^{+\infty}\partial_2^{-j}V_j\partial_1=-\sum_{j=1}^{\infty}
 \partial_2^{-j}V_j\partial_2^{-1}\rho+\rho^{-1}\partial_1V_1\partial_2^{-1}\rho,
\end{align*}
implying 
$$
  \sum_{j=1}^{\infty}
 \partial_2^{-j}V_j(\partial_1+\partial_2^{-1}\rho)=\rho^{-1}\partial_1V_1\partial_2^{-1}\rho.
$$
Since $V_1=-\partial_1^{-1}\rho\partial_1^{-1}$ \big(see (\ref{Vjprese1})\big) and $\sum_{j=1}^{+\infty}\partial_2^{-j}V_j=\iota_{\partial_2^{-1}}H^{-1}\partial_2-\partial_1^{-1}$, we can obtain
\begin{align*}
  (\iota_{\partial_2^{-1}}H^{-1}\partial_2-\partial_1^{-1})(\partial_1+\partial_2^{-1}\rho)=\partial_1^{-1}\partial_2^{-1}\rho,
\end{align*}  
which is just $
  H=\partial_1\partial_2+\rho.$ Therefore, $\pi_{1}(\partial_2^{k+1})=A_{k+1}$ holds.
\end{proof}
\end{proposition}
\subsection{From the bilinear equation to the Lax equation}\label{bilineartolax}
In this subsection, starting from the 2-BKP bilinear equation (\ref{2bkptaubilinear}), we will derive the corresponding Lax equation for the 2-BKP hierarchy, which contains the following two key steps:
\begin{itemize}
  \item from bilinear equation (\ref{2bkptaubilinear}) to the dressing operators,
  \item from the dressing operators to the Lax equation (\ref{2bkplaxtriple}).
\end{itemize}The first step has been discussed in \cite{Geng2023,Liu2011}, but for convenience, we will still contain this part to fix some notations.

For this, let us firstly introduce the 2-BKP wave function $\psi_a(\mathbf{t},z)$ by the way below:
\begin{align}\label{wavefunction}
&\psi_a(\mathbf{t},z)
=\frac{\tau(\mathbf{t}-2[z^{-1}]_a)}{\tau(\mathbf{t})}e^{\xi(t^{(a)},z)},\quad a=1,2.
\end{align}
Then (\ref{2bkptaubilinear}) will become,
\begin{align}
        {\rm Res}_z z^{-1}\psi_1(\mathbf{t},z)\psi_1(\mathbf{t}',-z)={\rm Res}_z z^{-1}\psi_2(\mathbf{t},z)\psi_2(\mathbf{t}',-z).\label{2bkptaubilinear2}
\end{align}
Next we define the dressing operators $W_a(\mathbf{t},\partial_{a})=1+\sum_{i=1}^{\infty}w_{a,i}(\mathbf{t})\partial_{a}^{-i}$ such that
\begin{align}\label{wavedress}
&\psi_a(\mathbf{t},z)
=W_a(\mathbf{t},\partial_{a})(e^{\xi(t^{(a)},z)}),\quad a=1,2.
\end{align}
Before further discussion, let us see the lemmas below.
\begin{lemma}\cite{Cheng2021-2,Date1983}\label{lemma:ref5lemma4}
Given two pseudo-differential operators $A(x,\partial_x)=\Sigma_ia_i(x)\partial_x^{i}$ and $B(x',\partial_{x'})=\Sigma_jb_j(x')\partial_{x'}^{j}$,
\begin{align*}
{\rm Res}_{z}A(x,\partial_x)(e^{xz})B(x',\partial_{x'})(e^{-x'z})=\big(A(x,\partial_x)B^*(x,\partial_x)
\partial_x\big)(\Delta^0),
\end{align*}
where $B^*(x,\partial_x)=\Sigma_j(-\partial_x)^{j}b_j(x), \Delta^0=(x-x')^0$, and
$\partial_x^{-a}\big((x-x')^0\big)=\frac{(x-x')^a}{a!}$ for $a\geq0$ and others are zero.
\end{lemma}
\begin{remark}\label{remarkaxbx}
  Notice that if $A(x,\partial_x)(\Delta^0)=B(x,\partial_x)(\Delta^0)$, then $A(x,\partial_x)_{\partial_x,\leq0}=B(x,\partial_x)_{\partial_x,\leq0}$.
\end{remark}
\begin{lemma}\cite{Wu2023}\label{dressinglemma}
Given two functions $f(x)$ and $g(x)$,
  \begin{align*}
    f(x)g(x')=(f(x)\partial_x^{-1}g(x)\partial_x)(\Delta^0).
  \end{align*}
\end{lemma}
\begin{proposition}\label{laxsatoprimelyprop}
The dressing operators $W_a(\mathbf{t},\partial_{a})(a=1,2)$ satisfy
\begin{align}
&W_a\partial_a^{-1}W^*_a\partial_a=1,\label{waeq2}\\ 
&\partial_{a,n}W_{a}=-(W_a\partial_a^nW_a^{-1})_{a,<0}W_a,\label{waeq1}\\ &\partial_{a,n}W_{3-a}=(W_a\partial_a^n\partial_{3-a}^{-1}W_a^{-1}
\partial_{3-a})_{a,[0]}W_{3-a}.\label{2mbkpsatoeq1}
\end{align}
\begin{proof}
Firstly by (\ref{wavefunction}) and Lemma \ref{lemma:ref5lemma4}, we have
\begin{align}\label{2bkpbil-reduction1}
\big(W_1(x_1,x_2)\partial_1^{-1}W_1^*(x_1,x'_2)\partial_1\big)\big((x_1-x'_1)^0\big)=\big(W_2(x_1,x_2)
\partial_2^{-1}W_2^*(x'_1,x_2)\partial_2\big)
\big((x_2-x'_2)^0\big),
\end{align}
where $W_a(x_1,x_2)$ means $W_a(\mathbf{t})$,  and all $t_k^{(a)}$ are the same expect $t_1^{(1)}=x_1$, $t_1^{(2)}=x_2$. Then if set $x'_1=x_1$ or $x'_2=x_2$ in (\ref{2bkpbil-reduction1}), then we can get (\ref{waeq2}) by Remark \ref{remarkaxbx}.

If apply $\partial_{1,n}$ to (\ref{2bkptaubilinear2}) and set $\mathbf{t'}=\mathbf{t}$ except $x'_1$ and $x'_2$ , then we have 
\begin{equation*}
    \begin{aligned}
        &\big((\partial_{1,n}W_1(x_1,x_2)+W_1(x_1,x_2)\partial_1^n)W_1(x_1,x'_2)^{-1}\big)_{1, <0}\big((x_1-x'_1)^0)
    \\&=\big(W_2(x_1,x_2)_{t^{(1)}_n}W_2(x'_1,x_2)^{-1}\big)\big((x_2-x'_2)^0\big).
    \end{aligned}
\end{equation*}
If set $x'_2=x_2$, we can obtain (\ref{waeq1}) for $a=1$, while if set $x'_1=x_1$, (\ref{2mbkpsatoeq1}) for $a=1$ can be obtained by Lemma \ref{dressinglemma}.
\end{proof}
\end{proposition}

\begin{corollary}\label{Hexist}
  There exists one operator $H=\partial_1\partial_2+\rho$ such that 
    $H(\psi_1)=H(\psi_2)=0$
  and
  \begin{align}\label{hweq}
     H=\partial_1W_1\partial_{2}W_1^{-1}=\partial_2W_2\partial_{1}W_2^{-1},
  \end{align}
 where $\rho=2\partial_1\partial_2(\log\tau)$.
 \begin{proof}
 If set $n=1$ in (\ref{2mbkpsatoeq1}), we have
\begin{align*}
\partial_a(W_{3-a})=-\partial_{3-a}^{-1}\rho\cdot W_{3-a},
\end{align*}
which means $HW_a=\partial_aW_a\partial_{3-a}(a=1,2)$, thus (\ref{hweq}) is correct. While $H(\psi_a)=0$ can be easily obtain by (\ref{hweq}) and (\ref{wavedress}).
 \end{proof}
\end{corollary}
By (\ref{hweq}), we can find $\iota_{\partial_a^{-1}}H^{-1}\partial_{3-a}=W_a\partial_{3-a}^{-1}W_a^{-1}$. Thus, we have the corollary below.
\begin{corollary}\label{Hwrelation}
  For $k\geq1$,
  \begin{align*}
\pi_{a}(\partial_{3-a}^k)=(\partial_{3-a}^kW_{3-a}\partial_a^{-1}W_{3-a}^{-1})_{3-a,[0]}\partial_a,\quad a=1,2.
\end{align*}
\end{corollary}
\begin{proposition}\label{relationdress}
$\partial_{a,n}\psi_b=B_n^{(a)}(\psi_{b})$ with $B_n^{(a)}=(W_a\partial_a^nW^{-1}_a)_{a,\geq0}$.
\begin{proof}
  Firstly, the case of $a=b$ can be easily proved by (\ref{waeq1}) and $\psi_a(\mathbf{t},z)
=W_a(\mathbf{t},\partial_{a})(e^{\xi(t^{(a)},z)})$.
Next by Corollary \ref{Hwrelation}, we know
\begin{align}\label{2mbkpsatoeq3}
\partial_{a,n}W_{b}=\pi_{b}\big(B_n^{(a)}\big)W_{b}, \quad a\neq b,
\end{align}
which implies $\partial_{a,n}\psi_{b}=\pi_b(B_{n}^{(a)})(\psi_b)(a\neq b)$. Further by $H(\psi_a)=0$, we can know $\partial_{a,n}\psi_{b}=B_{n}^{(a)}(\psi_b)$$\big(a\neq b\big)$.
\end{proof}
\end{proposition}

After the proposition above, now we can obtain the Lax equation of the 2-BKP hierarchy, which is given by the proposition below.
\begin{proposition}\label{Laxeqderive} If denote $L_a=W_a\partial_aW_a^{-1}$, and $B_n^{(a)}=(L_a^n)_{a,\geq0}$, then the Lax equation of the 2-BKP hierarchy is given by the Lax triple $(L_1,L_2,H)$ satisfying
\begin{align*}
&L_a^*=-\partial_aL_a\partial_a^{-1},\\
&\partial_{a,n}L_b=[\pi_{b}(B_n^{(a)}),L_{b}],\quad a,b=1,2,\\
  &\partial_{a,n}H=-(B^{(a)}_{n})^*H-HB^{(a)}_{n},
\end{align*}
where $\pi_{a}(\partial_{3-a}^k)=(\partial_{3-a}^k\iota_{\partial_a^{-1}}H^{-1}\partial_{3-a})_{3-a,[0]}\partial_a$.
\begin{proof}
Firstly, $L_a^*=-\partial_aL_a\partial_a^{-1}$ can be easily obtained by $W_a\partial_a^{-1}W^*_a=\partial_a^{-1}$. 
Next by (\ref{waeq1}) and (\ref{2mbkpsatoeq3}), we know
\begin{align}\label{dressingoper}
  \partial_{a,n}W_a=B_n^{(a)}W_a-W_a\partial_a^n,\quad 
    \partial_{a,n}W_b=\pi_b(B_n^{(a)})W_b,\quad a\neq b,
\end{align}
which gives rise to $\partial_{a,n}L_b$. As for $\partial_{a,n}H$, it can be obtained by $H=\partial_1W_1\partial_{2}W_1^{-1}=\partial_2W_2\partial_{1}W_2^{-1}$ and (\ref{dressingoper}).
\end{proof}
\end{proposition}
\begin{example}
Firstly by $\partial_{a}L_{3-a}=[-\partial_{3-a}^{-1}\rho,L_{a}]$, we can know
\begin{align*}
  \partial_2(u_{1,1})=\partial_1(\rho),\quad \partial_1(u_{2,1})=\partial_2(\rho).
\end{align*}
Further by $B_3^{(a)}=\partial_a^3+3u_{a,1}\partial_a$, $\partial_{a,3}H=-(B^{(a)}_{3})^*H-HB^{(a)}_{3}$ will give rise to
 \begin{align*}
 \partial_{a,3}\rho=\partial_a^3(\rho)+3\partial_a(u_{a,1}\rho),\quad a=1,2.
 \end{align*}
 Finally, if we take $t=-t^{(1)}_{3}-t^{(2)}_{3}$, $\widetilde{t}=-t^{(1)}_{3}+t^{(2)}_{3}$, $x_1=x$, $x_2=y$, then we can get the Novikov-Veselov equation \cite{Novikov1986,Hu1996},
  \begin{align*}
  &2\partial_{t}\rho+\rho_{xxx}+\rho_{yyy}+3(\rho\partial_y^{-1}\rho_x)_x
  +3(\rho\partial_x^{-1}\rho_y)_y=0.
\end{align*}
\end{example}

\subsection{From the Lax equation to the bilinear equation}\label{laxtobilinear}
Given the 2-BKP Lax operators $L_a=\partial_a+\sum_{i=1}^{+\infty}u_{a,i}\partial_a^{-i}$ ($a=1,2$) and a special operator $H=\partial_1\partial_2+\rho$, 
the Lax equation of the 2-BKP hierarchy is defined by a Lax triple $(L_1,L_2,H)$ satisfying
\begin{align}
 &L^*_a=-\partial_aL_a\partial_a^{-1},\label{LAXshow1}\\
  &\partial_{a,n}L_b=[\pi_{b}(B_n^{(a)}),L_{b}],\quad a,b=1,2,\label{LAXshow2}\\
  &\partial_{a,n}H=-(B^{(a)}_{n})^*H-HB^{(a)}_{n}.\label{LAXshow3}
\end{align}
Here $\pi_a$ is the projection: $\mathcal{E}_{(a)}=\mathcal{E}^0_{(a)}\oplus\mathcal{E}_{(a)}H\rightarrow\mathcal{E}_{(a)}^0$, which can be computed by $\pi_{a}(\partial_{3-a}^k)=(\partial_{3-a}^k\iota_{\partial_a^{-1}}H^{-1}\partial_{3-a})_{3-a,[0]}\partial_a$.

Firstly, the system of (\ref{LAXshow1})--(\ref{LAXshow3}) is well defined, due to the following two lemmas.
\begin{lemma}\label{Hla}
  $H L_a=\partial_a L_a\partial_a^{-1}H=-L_a^*H$, $a=1,2$.
  \begin{proof}
    If we set $a=2,b=1,n=1$ in (\ref{LAXshow2}), we have
\begin{align*}
  \partial_2(L_1)&=[\pi_1(B_1^{(2)}),L_1]=[\pi_1(\partial_2),L_1]\\&
  =[-\partial_1^{-1}\rho,L_1]=-\partial_1^{-1}\rho L_1+L_1\partial_1^{-1}\rho\\&
  =-(\partial_1^{-1}H-\partial_2)L_1+L_1(\partial_1^{-1}H-\partial_2),
\end{align*}
then $H L_1=\partial_1 L_1\partial_1^{-1}H$. Similarly by $a=1,b=2,n=1$ in (\ref{LAXshow2}), we can derive
$H L_2=\partial_2 L_2\partial_2^{-1}H$.
  \end{proof}
\end{lemma}
\begin{lemma}\label{welldefined}
  In the 2-BKP hierarchy,
  \begin{align*}
    \pi_a(B_m^{(3-a)})^*=(-1)^{m}\pa_a\cdot  \pi_a(B_m^{(3-a)})\cdot \pa_a^{-1},\quad a=1,2,\quad m\in\mathbb{Z}_{>0}.
  \end{align*}
  \begin{proof}
   Firstly, let us consider $a=2$. Starting from
    \begin{align*}
  \pi_{2}(B_m^{(1)})&={\rm Res}_{\partial_1}B_{m}^{(1)}\cdot\iota_{\partial_2^{-1}}H^{-1}\partial_2={\rm Res}_{\partial_1}L_1^m\cdot\iota_{\partial_2^{-1}}H^{-1}\partial_2,
\end{align*}
we have:
\begin{align*}
  \pi_{2}(B_m^{(1)})^*&=-{\rm Res}_{\partial_1}(L_1^m\cdot\iota_{\partial_2^{-1}}H^{-1}\partial_2)^*\\&={\rm Res}_{\partial_1}\partial_2\cdot\iota_{\partial_2^{-1}}H^{-1}(L_1^m)^*
 \\&=(-1)^m\partial_2\cdot{\rm Res}_{\partial_1} L_1^n\cdot\iota_{\partial_2^{-1}}H^{-1}\partial_2\cdot\partial_2^{-1}
 \\&=(-1)^m\partial_2\cdot\pi_{2}(B_m^{(1)})\cdot\partial_2^{-1}.
\end{align*}
Here in the second equality, ${\rm Res}_{\partial_1}A^*=-{\rm Res}_{\partial_1}A$ is used. Further in the third equality, Lemma \ref{Hla} is used. The case when $a=1$ can be derived in the same way. 
  \end{proof} 
\end{lemma}
\begin{remark}
By Lemma \ref{welldefined}, we can know $L^*_a=-\partial_aL_a\partial_a^{-1}$ can be kept under $\partial_{b,n}$. Thus the system (\ref{LAXshow1})--(\ref{LAXshow3}) is well-defined. 
\end{remark}

Next, let us discuss the commutativity of flows, which is given in the proposition below.
\begin{proposition}\label{flowsex}
Given odd integers $m,n\geq0$ and $a,b,c=1,2,$
\begin{align}\label{flowex1}
\partial_{a,m}B^{(b)}_{n}-\partial_{b,n}B^{(a)}_{m}+[B^{(b)}_{n},B^{(a)}_{m}]\in \mathcal{E} H,
\end{align}
where $\mathcal{E}=\mathcal{B}[\partial_1,\partial_2]$. Thus, we have
\begin{align}\label{flowex3}
  \partial_{a,m}\pi_c(B^{(b)}_{n})-\partial_{b,n}\pi_c(B^{(a)}_{m})+[\pi_c(B^{(b)}_{n}),\pi_c(B^{(a)}_{m})]=0.
\end{align}
And further 
   $[\partial_{a,m},\partial_{b,n}]L_c=0,$ $[\partial_{a,m},\partial_{b,n}]H=0.$ 
\begin{proof}
Firstly by direct computation, one can easily find
$\partial_{a,m}B^{(a)}_{n}-\partial_{a,n}B^{(a)}_{m}+[B^{(a)}_{n},B^{(a)}_{m}]=0$. So (\ref{flowex1}) is correct when $a=b$.

 Next let us prove (\ref{flowex1}) for the case of $a=1$, $b=2$. Actually by using $H=\partial_1\partial_2+\rho$, we can rewrite
\begin{align}\label{flowexchange1}
[B^{(2)}_{n},B^{(1)}_{m}]=h_{mn}+\pi_2([B^{(2)}_{n},B^{(1)}_{m}])_{2,\geq1}+\pi_1([B^{(2)}_{n},B^{(1)}_{m}])_{1,\geq1}+E_{mn}H,
\end{align}
where $E_{mn}\in \mathcal{E}$. Notice that by $HL_a=\partial_aL_a\partial_a^{-1}H$ in Lemma \ref{Hla} and (\ref{LAXshow2}), we can know $\partial_{1,m}B^{(2)}_{n}=\pi_2([B^{(1)}_{m},L^{n}_{2}])_{2,\geq1}.$
Further by Lemma \ref{piaaction}, we have $\pi_2([B^{(1)}_{m},(L^{n}_{2})_{2,<0}])_{2,\geq1}=0$. Thus 
\begin{align}\label{flowexchange2}
\partial_{1,m}B^{(2)}_{n}=\pi_2([B^{(1)}_{m},B^{(2)}_{n}])_{2,\geq1}.
\end{align}
Similarly, we can get 
\begin{align}\label{flowexchange3}
 \partial_{2,n}B^{(1)}_{m}=\pi_1([B^{(2)}_{n},B^{(1)}_{m}])_{1,\geq1}.
\end{align}
Therefore by (\ref{flowexchange1})--(\ref{flowexchange3}), we only need to show $h_{mn}=0$. Indeed,
\begin{align*}
h_{mn}=\pi_1([B^{(1)}_{m},B^{(2)}_{n}])_{2,[0]}=-(\partial_{2,n}L_1^m)_{1,[0]}=0.
\end{align*}
So we have proved (\ref{flowex1}) for $a=1$, $b=2$. And (\ref{flowex1}) for $a=2$, $b=1$ can be similarly proved.

Based on (\ref{flowex1}), we can get (\ref{flowex3}) by (\ref{LAXshow3}). And $[\partial_{a,m},\partial_{b,n}]L_c=0$, $[\partial_{a,m},\partial_{b,n}]H=0$ can be proved directly by (\ref{LAXshow2}) and (\ref{LAXshow3}).
\end{proof}
\end{proposition}
\begin{lemma}\label{laxtodress}
  Given $L=\partial_x+\sum_{i=1}^{+\infty}u_{i+1}\partial_x^{-i}$ satisfying $L^*=-\partial_x L\partial_x^{-1}$, there exists $W=1+\sum_{j=1}^{\infty}w_{j}\partial_x^{-j}$ such that $L=W\partial_x W^{-1}$ and $W\partial_x^{-1}W^*=\partial_x^{-1}$.
\begin{proof}
From $LW=W\partial_x$, we can know 
\begin{align*}
 u_2=-w_{1,x},\quad u_{i+1}=-w_{i,x}+p_i(w_1,w_2,\ldots,w_{i-1}),\quad i\geq2,
\end{align*}
where $p_i$ is the differential polynomial of $w_1,w_2,\ldots,w_{i-1}$. Thus we can express $w_i$ in terms of $u_j$ (2$\leq j\leq i+1$), which means there exists $\widetilde{W}=1+\sum_{i=1}^{\infty}\widetilde{w}_i\partial_x^{-i}$ such that $L=\widetilde{W}\partial_x \widetilde{W}^{-1}$.

Next by  $L^*=-\partial_x L\partial_x^{-1}$, we can know 
\begin{align*}
  [\widetilde{W}^*\partial_x\widetilde{W},\partial_x]=0.
\end{align*}
Thus if denote  $C=\widetilde{W}^*\partial_x\widetilde{W}=\partial_x+\sum_{i\geq0}c_i\partial_x^{-i}$, then $c_{i,x}=0$. Notice that $C^*=-C$, so $C$ can be further written into
\[C=\partial_x+\sum_{i=1}^{\infty}c_{2i-1}\partial_x^{-2i+1}.
\]
It can be proved that there exists $\widetilde{C}=1+\sum_{i=1}^{+\infty}\widetilde{c}_{2i}\partial_x^{-2i}$ such that $C=\widetilde{C}^*\partial_x\widetilde{C}$. In fact by the direct computation, we have
\[
c_1=2\widetilde{c}_2,\quad c_{2i-1}=2\widetilde{c}_{2i}+\sum_{j=1}^{i-1}\widetilde{c}_{2(i-j)}\widetilde{c}_{2j},\quad i\geq2.
\]
So we can easily express $\widetilde{c}_{2i}$ by $c_{2j-1}(1\leq j\leq i)$. After the preparation above, if we set $W=\widetilde{W}\widetilde{C}^{-1}$, then $W\partial_x W^{-1}=\widetilde{W}\partial_x \widetilde{W}^{-1}=L$ and
\[W\partial_x^{-1} W^*=\widetilde{W}\widetilde{C}^{-1}\partial_x^{-1}\widetilde{C}^{*-1}\widetilde{W}^*=\widetilde{W}C^{-1}\widetilde{W}^*
=\partial_x^{-1}.\]
\end{proof}
\end{lemma}
\begin{proposition}\label{wavefunctionde}
Given the 2-BKP Lax triple $(L_1,L_2,H)$ with $L_a=\partial_a+\sum_{i=1}^{+\infty}u_{a,i}\partial_a^{-i}$ and $H=\partial_1\partial_2+\rho$ satisfying
\begin{align*}
&L_a^*=-\partial_aL_a\partial_a^{-1},\\
&\partial_{a,n}L_b=[\pi_{b}(B_n^{(a)}),L_{b}],\quad a,b=1,2,\\
  &\partial_{a,n}H=-(B^{(a)}_{n})^*H-HB^{(a)}_{n},
\end{align*}
 then there exists $W_a=1+\sum_{i=1}^{\infty}w_{a,i}\partial_{a}^{-i}$ such that $L_a=W_a\partial_aW_a^{-1}$ satisfying 
\begin{align*}
&W_a\partial_a^{-1}W^*_a\partial_a=1,\quad H=\partial_1W_1\partial_{2}W_1^{-1}=\partial_2W_2\partial_{1}W_2^{-1},\quad a=1,2,\\
&\partial_{a,n}W_{a}=-(W_a\partial_a^nW_a^{-1})_{a,<0}W_a,\quad \partial_{a,n}W_{3-a}=(W_a\partial_a^n\partial_{3-a}^{-1}W_a^{-1}\partial_{3-a})_{a,[0]}W_{3-a}.
\end{align*}
\begin{proof} Firstly by Lemma \ref{laxtodress}, there exists $\widetilde{W}_a=1+\sum_{i=1}^{\infty}w_{a,i}\partial_{a}^{-i}$ satisfying $L_a=\widetilde{W}_a\partial_a\widetilde{W}_a^{-1}$ and $\widetilde{W}_a\partial_a^{-1}\widetilde{W}^*_a=\partial_a^{-1}$.
 Next, we consider the system below: 
   \begin{align}\label{Wchu}
\begin{split}
\left \{
\begin{array}{ll}
      \partial_{a,n}W_{a}=B_n^{(a)}W_a-L_a^nW_a,\\
      \partial_{a,n}W_{b}=\pi_{b}\big(B_n^{(a)}\big)W_{b}, \quad a\neq b,\\
      W_a|_{\mathbf{t}=0}=\widetilde{W}_a(0),\quad a,b=1,2,
\end{array}
\right.
\end{split}
\end{align}
where $B_n^{(a)}=(L_a^{n})_{a,\geq0}$ and $W_a$ has the form $1+\sum_{i=1}^{\infty}w_{a,i}\partial_{a}^{-i}$. By (\ref{flowex3}), the system (\ref{Wchu}) has a unique solution $\widehat{W}_a=1+\sum_{i=1}^{\infty}\widehat{w}_{a,i}\partial_{a}^{-i}$. Then by (\ref{flowex3}), we have
\begin{align*}
  &\partial_{a,n}(L_{a}\widehat{W}_a-\widehat{W}_a\partial_a)=B_n^{(a)}(L_{a}\widehat{W}_a-\widehat{W}_a\partial_a)-L_a^n(L_{a}\widehat{W}_a-\widehat{W}_a\partial_a),\\
  &\partial_{a,n}(L_{b}\widehat{W}_b-\widehat{W}_b\partial_b)=\pi_{b}\big(B_n^{(a)}\big)(L_{b}\widehat{W}_b-\widehat{W}_b\partial_b),\quad a\neq b,
  \end{align*}
   and $(L_{a}\widehat{W}_a-\widehat{W}_a\partial_a)|_{\mathbf{t}=0}=0.$ So we can find
   $\widehat{V}_a=L_{a}\widehat{W}_a-\widehat{W}_a\partial_a$ is the solution of the following system:
   \begin{align}\label{Wchu2}
\begin{split}
\left \{
\begin{array}{ll}
      \partial_{a,n}V_{a}=B_n^{(a)}V_a-L_a^nV_a,\\
      \partial_{a,n}V_{b}=\pi_{b}\big(B_n^{(a)}\big)V_{b}, \quad a\neq b,\\
      V_a|_{\mathbf{t}=0}=0,\quad a,b=1,2,
\end{array}
\right.
\end{split}
\end{align}
  where the unknown operator $V_a$ has the form $\sum_{j=1}^{\infty}v_{a,j}\partial_a^{-j}$. Similarly by (\ref{flowex3}), we can find (\ref{Wchu2}) also has the unique solution, while $V_a=0$ is one solution of (\ref{Wchu2}), thus $\widehat{V}_a=0$, which means $L_a=\widehat{W}\partial_a\widehat{W}^{-1}$.

  If we set $n=1$ in $\partial_{1,1}\widehat{W}_{2}=\pi_{2}\big(B_1^{(1)}\big)\widehat{W}_{2}$, then 
$
  \partial_1(\widehat{W}_{2})=-\partial_{2}^{-1}\rho\cdot \widehat{W}_{2},$
  which is $H=\partial_2\widehat{W}_2\partial_{1}\widehat{W}_2^{-1}.$ Similarly we can get $H=\partial_1\widehat{W}_1\partial_{2}\widehat{W}_1^{-1}$. Based on this,  
  $\pi_{b}\big(B_n^{(a)}\big)=(\widehat{W}_a\partial_a^n\partial_{b}^{-1}\widehat{W}_a^{-1}\partial_{b})_{a,[0]}(a\neq b)$, so $\partial_{a,n}\widehat{W}_b=(\widehat{W}_a\partial_a^n\partial_{b}^{-1}\widehat{W}_a^{-1}\partial_{b})_{a,[0]}\widehat{W}_b$ ($a\neq b$).
\end{proof}
  \end{proposition}
  \begin{remark}
    In fact, $W_1$ and $W_2$ can be up to the right multiplication by the constant operators $C_1=\partial_{1}+\sum_{i=1}^{+\infty}c_{1,i}\partial_{1}^{-i}$ and  $C_2=\partial_{2}+\sum_{i=1}^{+\infty}c_{2,i}\partial_{2}^{-i}$ respectively. Here the coefficients $c_{a,i}$ ($a=1,2$) are independent of $x_a$. 
  \end{remark}

If we introduce the 2-BKP wave function $\psi_a(\mathbf{t},z)$ by
$$\psi_a(\mathbf{t},z)
=W_a(\mathbf{t},\partial_{a})(e^{\xi(t^{(a)},z)}),$$
then 
$\partial_{a,n}\psi_b(\mathbf{t},z)=B_n^{(a)}\big(\psi_{b}(\mathbf{t},z)\big)$. And by $H=\partial_1W_1\partial_{2}W_1^{-1}=\partial_2W_2\partial_{1}W_2^{-1}$, we can know
\begin{align}\label{hpsi0}
 H\big(\psi_{a}(\mathbf{t},z)\big)=0, \quad a=1,2.
\end{align}
Further when $a\neq b$ for $a,b=1,2$, $\partial_{a,n}W_{b}=\pi_{b}\big(B_n^{(a)}\big)\cdot W_{b}$. So we have  $\partial_{a,n}\psi_{b}=\pi_b(B_{n}^{(a)})(\psi_b)$$\big(a\neq b\big)$. Then by 
$B_n^{(a)}-\pi_b(B_{n}^{(a)})\in \mathcal{E}_{(b)}H$ and (\ref{hpsi0}), we can find 
$\partial_{a,n}\psi_b=B_n^{(a)}(\psi_{b})\big(a\neq b\big)$, which implies that
\begin{align*}
\partial_{a,n}W_{b}=B_n^{(a)}\big(W_{b}\big), \quad a\neq b.
\end{align*}
Here $B_n^{(a)}\big(W_{b}\big)$ means $B_n^{(a)}$ acting on coefficients of $W_{b}$.

For the 2-BKP dressing operators $W_1$ and $W_2$, we have found that they satisfy the following relations:
  \begin{align*}
&W_a\partial_a^{-1}W^*_a\partial_a=1,\quad \partial_1W_1\partial_{2}W_1^{-1}=\partial_2W_2\partial_{1}W_2^{-1},\quad a=1,2,\\ 
&\partial_{a,n}W_{a}=-(W_a\partial_a^nW_a^{-1})_{a,<0}W_a,\quad \partial_{a,n}W_{3-a}=(W_a\partial_a^nW_a^{-1})_{a,\geq0}(W_{3-a}).
\end{align*}
Thus by Theorem 3.9 in \cite{Geng2023}, we can obtain the bilinear equation of the 2-BKP hierarchy in terms of the wave function, that is,
\begin{align}\label{bilineareq}
     {\rm Res}_z z^{-1}\psi_1(\mathbf{t},z)\psi_1(\mathbf{t'},-z)={\rm Res}_z z^{-1}\psi_2(\mathbf{t},z)\psi_2(\mathbf{t'},-z).
  \end{align}   
Further by Proposition 3.15 in \cite{Geng2023}, there exists the 2-BKP tau function $\tau(\mathbf{t})$ such that
\begin{align}\label{psitaurelation}
\psi_a(\mathbf{t},z)=\frac{\tau(\mathbf{t}-2[z^{-1}]_a)}{\tau(\mathbf{t})}e^{\xi(t^{(a)},z)},\quad a=1,2.
\end{align}
After substituting (\ref{psitaurelation}) into (\ref{bilineareq}), we can finally obtain the 2-BKP bilinear equation (\ref{2bkptaubilinear}) in terms of the tau function.

By now we have showed the equivalence of the Lax equation and the bilinear equation in the case of the 2-BKP hierarchy.

\subsection{The $(M_1,M_2)$-reduction of the 2-BKP hierarchy}\label{reductions2bkp}
In this subsection, let us define the $(M_1,M_2)$-reduction of the 2-BKP hierarchy, which is given by the following constraints on the 2-BKP Lax operators $L_1$ and $L_2$:
\begin{align}\label{defineconstranit}
  (L_a^{M_a})_{a,\leq0}=\pi_a(B_{M_{3-a}}^{(3-a)}),\quad M_1+M_2\text{ is even}, \quad a=1,2.
\end{align}
Since $\big(L_a^n\big)_{a,[0]}=0$ for odd $n$ and $\big(L_a^m\big)_{a,[0]}\neq0$ for even $m$ (see \cite{Date1983}), thus if $M_1+M_2$ is odd, then we find the highest order terms of (\ref{defineconstranit}) are not consistent. So $M_1+M_2$ can only be even.

Firstly, let us see the constraints (\ref{defineconstranit}) are well defined, which is given in the proposition below.
\begin{proposition}
  For the $(M_1,M_2)$-reduction (\ref{defineconstranit}) of the 2-BKP hierarchy,
  \begin{align*}
    \pa_{b,n}L_a^{M_a}=\pa_{b,n}\big(\widetilde{B}_{M_a}^{(a)}+\pi_a(B_{M_{3-a}}^{(3-a)})\big),\quad
    a,b=1,2,
  \end{align*}
  where $\widetilde{B}_{M_a}^{(a)}=(L_a^{M_a})_{a,\geq1}$, $B_{M_a}^{(a)}=(L_a^{M_a})_{a,\geq0}$.
  \begin{proof}
  Firstly, let us prove the case for $a=b=1$ and denote 
  \[A=\pa_{1,n}\big(\widetilde{B}_{M_1}^{(1)}+\pi_1(B_{M_2}^{(2)})\big)-\pa_{1,n}L_1^{M_1}.\]
  Notice that $A\in\E_1^{(0)}$, so by Proposition \ref{proH1}, if we can prove $A\in\E_1H$, then we will get $A = 0$.

  In fact, by $\pa_{1,n}L_1=[B_n^{(1)},L_1]$, we can know
  \begin{align*}
    A&=[B_{n}^{(1)},L_1^{M_1}]_{1,\geq1}+\pa_{1,n}\pi_1(B_{M_2}^{(2)})-[B_{n}^{(1)},L_1^{M_1}].
  \end{align*}
  Next by $L_1^{M_1}=\big(L_1^{M_1}\big)_{1,\geq1}+\big(L_1^{M_1}\big)_{1,\leq0}=\widetilde{B}_{M_1}^{(1)}+\pi_1(B_{M_2}^{(2)})$, where (\ref{defineconstranit}) is used, and the relation $[B_{n}^{(1)},\widetilde{B}_{M_1}^{(1)}]_{1,\geq1}=[B_{n}^{(1)},\widetilde{B}_{M_1}^{(1)}]$, we have
  \begin{align*}
    A&=[B_{n}^{(1)},\pi_1(B_{M_2}^{(2)})]_{1,\geq1}+\pa_{1,n}\pi_1(B_{M_2}^{(2)})-[B_{n}^{(1)},\pi_1(B_{M_2}^{(2)})]
    \\&=[L_1^n,\pi_1(B_{M_2}^{(2)})]_{1,\geq1}+\pa_{1,n}\pi_1(B_{M_2}^{(2)})-[B_{n}^{(1)},\pi_1(B_{M_2}^{(2)})],
  \end{align*}
  where we have used $[\big(L_1^n\big)_{1,\leq-1},\pi_1(B_{M_2}^{(2)})]_{1,\geq1}=0$. If assume $\pi_1(B_{M_2}^{(2)})=B_{M_2}^{(2)}-A_1H$ for $A_1\in\E_1$, then 
 \begin{align*}
    A=&[L_1^n,\pi_1(B_{M_2}^{(2)})]_{1,\geq1}
    +\pa_{1,n}\big(B_{M_2}^{(2)}-A_1H\big)-[B_{n}^{(1)},B_{M_2}^{(2)}-A_1H]\\
    =&[L_1^n,\pi_1(B_{M_2}^{(2)})]_{1,\geq1}
    +[\pi_2(B_n^{(1)}),L_2^{M_2}]_{2,\geq0}-\pa_{1,n}(A_1H)-[B_{n}^{(1)},B_{M_2}^{(2)}-A_1H],
  \end{align*}
where $\pa_{1,n}L_2=[\pi_2(B_n^{(1)}),L_2]$ is used.

Next by similar method as (\ref{flowexchange1}), we know
\begin{align*}
  [B_{n}^{(1)},B_{M_2}^{(2)}]\in\pi_1([B_n^{(1)},B_{M_2}^{(2)}])_{1,\geq1}+\pi_2([B_n^{(1)},B_{M_2}^{(2)}])_{2,\geq0}+\E H,
\end{align*}
and by $HL_1=\pa_1L_1\pa_1^{-1}H$, we can get 
\begin{align*}
  [L_1^n,\pi_1(B_{M_2}^{(2)})]_{1,\geq1}=\pi_1([B_{n}^{(1)},B_{M_2}^{(2)}])_{1,\geq1},\quad 
  [\pi_2(B_{n}^{(1)}),L_{2}^{M_2})]_{2,\geq0}=\pi_2([B_{n}^{(1)},B_{M_2}^{(2)}])_{2,\geq0}.\end{align*}
  Based upon these facts, we can finally obtain
  \begin{align*}
    A\in-\pa_{1,n}(A_1H)+[B_{n}^{(1)},A_1H]+\E H.
\end{align*}
Furthermore, from the relation $\partial_{1,n}H=-(B^{(1)}_{n})^*H-HB^{(1)}_{n}$, we obtain $A \in \mathcal{E}_1 H$, while $A$ also belongs to $\mathcal{E}_1^{0}$. Thus, by Proposition \ref{proH1}, we can obtain $A=0$.

Other cases for $a\cdot b\neq 1$ can be similarly proved.
  \end{proof}
\end{proposition}

From (\ref{defineconstranit}), we can know
\begin{align}\label{constraints1}
(L_1^{M_1})_{1,[0]}=(L_2^{M_2})_{2,[0]}.
\end{align}
Thus for the $(M_1,M_2)$-reduction of the 2-BKP hierarchy, we can define the operator 
\begin{align}\label{mathcallexpress}
\mathcal{L}=\widetilde{B}_{M_1}^{(1)}+B_{M_2}^{(2)}, \quad a=1,2.
\end{align}  
Notice that $\mathcal{L}$ can be written into the forms below,
\begin{align}\label{mnreduction}
     \mathcal{L}=
\begin{split}
\left \{
\begin{array}{ll}
      \sum_{a=1}^{2}\Big(\pa_a^{M_a}+\sum_{l=1}^{\frac{M_a-3}{2}}(v_{a,l}\pa_a^{2l}
    +\pa_a^{2l}v_{a,l})\pa_a+v_{a,0}\pa_a\Big),\quad M_{a}~\rm{is~odd},\\
     \sum_{a=1}^{2}\Big(\pa_a^{M_a}+\sum_{l=1}^{\frac{M_a-2}{2}}(\widetilde{v}_{a,l}\pa_a^{2l-1}
    +\pa_a^{2l-1}\widetilde{v}_{a,l})\pa_a\Big)+\widetilde{v}_{0},\quad M_{a}~\rm{is~even}.
\end{array}
\right.
\end{split}
\end{align}
Then by (\ref{defineconstranit}), (\ref{constraints1}) and $H(\psi_a)=0$, we can easily obtain the proposition below.
\begin{proposition}\label{piallam}
  For the $(M_1,M_2)$-reduction (\ref{defineconstranit}) of the 2-BKP hierarchy,
  \begin{align*}
    \pi_a(\mathcal{L})=L_a^{M_a},
  \end{align*}
  and thus $\mathcal{L}(\psi_a)=z^{M_a}\psi_a$, $a=1,2$.
\end{proposition}
The next proposition gives an important equivalent formulation of the $(M_1,M_2)$-reduction of the 2-BKP hierarchy.
\begin{proposition}\label{reductionbilinear}
  For the $(M_1,M_2)$-reduction of the 2-BKP hierarchy, (\ref{defineconstranit}) is equivalent to
  \begin{align}\label{2mnbkptaubilinear}
   {\rm Res}_z  z^{M_1-1}\psi_1(\mathbf{t},z)\psi_1(\mathbf{t'},-z)={\rm Res}_z z^{M_2-1}\psi_2(\mathbf{t},z)\psi_2(\mathbf{t'},-z).
  \end{align}
  \begin{proof}
    Firstly starting from (\ref{defineconstranit}), we can prove (\ref{2mnbkptaubilinear}) by acting the operator $\mathcal{L}$ in (\ref{mnreduction}) on the 2-BKP bilinear equation
    \begin{align*}
        {\rm Res}_z z^{-1}\psi_1(\mathbf{t},z)\psi_1(\mathbf{t}',-z)={\rm Res}_z z^{-1}\psi_2(\mathbf{t},z)\psi_2(\mathbf{t}',-z),
\end{align*}
and using $\mathcal{L}(\psi_a)=z^{M_a}\psi_a$ in Proposition \ref{piallam}.

Then let us prove (\ref{defineconstranit}) from (\ref{2mnbkptaubilinear}). For this by
$\psi_a(\bft,z)=W_a(\bft,\pa_a)\big(e^{\xi(t^{(a)},z)}\big)$, $W_a\partial_a^{-1}W^*_a=\partial_a^{-1}$ and Lemma \ref{lemma:ref5lemma4}, we can know from (\ref{2mnbkptaubilinear}) that
\begin{align*}
  &\big(W_1(\bft,\pa_1)\partial_1^{M_1}W_1(\bft',\pa_1)^{-1}\big)\big((x_1-x'_1)^0\big)|
  _{\bft'=\bft~{\rm except}~
  x_2'\neq x_2}
  \\&=\big(W_2(\bft,\pa_2)\partial_2^{M_2}W_2(\bft',\pa_2)^{-1}\big)\big((x_2-x'_2)^0\big)
  _{\bft'=\bft~{\rm except}~
  x_1'\neq x_1}.
\end{align*}
If set $x_2'= x_2$, we can obtain by Lemma \ref{dressinglemma} that
\begin{align*}
  (L_1^{M_1})_{1,\leq0}=(L_2^{M_2}W_2\partial_1^{-1}W_2^{-1})_{2,[0]}\partial_1=
        \pi_1(B_{M_2}^{(2)}),
\end{align*}
where we have used Corollary \ref{Hwrelation}.
  \end{proof}
  By $\psi_a(\mathbf{t},z)
=\frac{\tau(\mathbf{t}-2[z^{-1}]_a)}{\tau(\mathbf{t})}e^{\xi(t^{(a)},z)}$ and (\ref{2mnbkptaubilinear}), we can get that the tau function of the $(M_1,M_2)$-reduction of the 2-BKP hierarchy satisfies
\begin{equation}
    \begin{aligned}
        &{\rm Res}_z z^{M_1-1}\tau(\mathbf{t}-2[z^{-1}]_1)\tau(\mathbf{t}'+2[z^{-1}]_1)e^{\xi(t^{(1)}-t^{(1)'},z)}
\\&={\rm Res}_z z^{M_2-1}\tau(\mathbf{t}-2[z^{-1}]_2)\tau(\mathbf{t}'+2[z^{-1}]_2)e^{\xi(t^{(2)}-t^{(2)'},z)},
\label{2m1m2bkptaubilinear}
    \end{aligned}
\end{equation}
 which is another equivalent formulation of the $(M_1,M_2)$-reduction (\ref{defineconstranit}).

\end{proposition}
For the Lax operator $\mathcal{L}$ of the $(M_1,M_2)$-reduction of the 2-BKP hierarchy, we have the following results. 
\begin{proposition}
    For the ($M_1,M_2$)-reduction of the 2-BKP hierarchy, 
    \begin{align}
        &H\mathcal{L}=(-1)^{M_a}\mathcal{L}^*H,\label{HLLH}\\
        &\pa_{a,n}\mathcal{L}-[B_n^{(a)},\mathcal{L}]\in \E H, \quad a=1,2\label{PATRALh}.
    \end{align}
    \begin{proof}
    Firstly, from (\ref{constraints1}) and (\ref{mnreduction}), we can know $(L_1^{M_1})_{1,[0]}=(L_2^{M_2})_{2,[0]}=v_0$. If we denote $\widetilde{B}_{M_1}^{(1)}=(L_1^{M_1})_{1,\geq1}$, $\widetilde{C}_{M_1}^{(1)}=(L_1^{M_1})_{1,\leq1}$, $B_{M_2}^{(2)}=(L_2^{M_2})_{2,\geq0}$, $C_{M_2}^{(2)}=(L_2^{M_2})_{2,\leq0}$,
        then by $L^*_a=-\partial_aL_a\partial_a^{-1}$, we can know
        \begin{equation}\label{BCjoint}
        \begin{aligned}
          &\big(\widetilde{B}_{M_1}^{(1)}\big)^*=(-1)^{M_1}\pa_2\cdot \widetilde{B}_{M_1}^{(1)}\cdot \pa_1^{-1},\quad \big(B_{M_2}^{(2)}\big)^*=(-1)^{M_2}\big(\pa_2\cdot B_{M_2}^{(2)}\cdot \pa_2^{-1}-\pa_2(v_0)\cdot \pa_2^{-1}\big),\\
          & \big(\widetilde{C}_{M_1}^{(1)}\big)^*=(-1)^{M_1}\pa_2\cdot \widetilde{C}_{M_1}^{(1)}\cdot \pa_1^{-1},\quad 
             \big(C_{M_2}^{(2)}\big)^*=(-1)^{M_2}\big(\pa_2\cdot C_{M_2}^{(2)}\cdot \pa_2^{-1}+\pa_2(v_0)\cdot \pa_2^{-1}\big).
        \end{aligned} 
        \end{equation}
        Thus by Lemma \ref{Hla}, we can know
        \begin{align*}
            HL_1^{M_1}=(-1)^{M_1}(L_1^{M_1})^*H,
        \end{align*}
        which means 
        \begin{align*}
            H\widetilde{B}_{M_1}^{(1)}-(-1)^{M_1}\big(\widetilde{B}_{M_1}^{(1)}\big)^*H
            =-H\widetilde{C}_{M_1}^{(1)}+(-1)^{M_1}\big(\widetilde{C}_{M_1}^{(1)}\big)^*H,
        \end{align*}
        and further
        \begin{align*}
            H\widetilde{B}_{M_1}^{(1)}-\pa_1\widetilde{B}_{M_1}^{(1)}\pa_1^{-1}H=\pa_1\widetilde{C}_{M_1}^{(1)}\pa_1^{-1}H-H\widetilde{C}_{M_1}^{(1)}.
        \end{align*}
       By comparing the coefficients of \( \pa_1^i \) on both sides, we can get
        \begin{align}\label{hl1leadsb2}
            H\widetilde{B}_{M_1}^{(1)}-\pa_1\widetilde{B}_{M_1}^{(1)}\pa_1^{-1}H=-\pa_1\pa_2(v_0)-\pa_2(v_0)\cdot\pa_1.
        \end{align}
        Similarly, we can obtain
        \begin{align}\label{hl1leadsb3}
            HB_{M_2}^{(2)}-\big(\pa_2B_{M_2}^{(2)}\pa_2^{-1}-\pa_2(c)\pa_2^{-1}\big)H
            =\pa_1\pa_2(v_0)+\pa_2(v_0)\cdot\pa_1.
        \end{align}
        Therefore, (\ref{hl1leadsb2}) and (\ref{hl1leadsb3}) mean that
        \begin{align*}
H\Big(\widetilde{B}_{M_1}^{(1)}+B_{M_2}^{(2)}\Big)-(-1)^{M_1}\Big(\big(\widetilde{B}_{M_1}^{(1)}\big)^*+\big(B_{M_2}^{(2)}\big)^*\Big)H=0,
        \end{align*}
        i.e., $H\mathcal{L}=(-1)^{M_a}\mathcal{L}^*H$, where we should note that $M_1+M_2$ is even.

       As for (\ref{PATRALh}), we can know from $\pa_{a,n}L_b=[\pi_b(B_n^{(a)}),L_b]$ and 
       $(L_1^{M_1})_{1,\leq0}=\pi_1(B_{M_2}^{(2)})$ that
       \begin{align*}
        \pa_{1,n}\mathcal{L}-[B_n^{(1)},\mathcal{L}]=\pi_1\big([B_n^{(1)},B_{M_2}^{(2)}]\big)_{1,\geq1}
        +\pi_2\big([B_n^{(1)},B_{M_2}^{(2)}]\big)
        _{2,\geq0}-[B_n^{(1)},B_{M_2}^{(2)}]
        \in\E H.
       \end{align*}
    \end{proof}
\end{proposition}
\begin{remark}\label{oddm1m2}
  When $M_1$ and $M_2$ are odd positive integers, we can find $\mathcal{L}=B_{M_1}^{(1)}+B_{M_2}^{(2)}$. So if we apply $\mathcal{L}^{k}(\bft,\pa_1,\pa_2)$ $(k\geq1, odd)$ to (\ref{2mnbkptaubilinear}) and use $\mathcal{L}(\psi_a)=z^{M_a}\psi_a$, then we can get 
  \begin{align*}
   {\rm Res}_z  z^{M_1k-1}\psi_1(\mathbf{t},z)\psi_1(\mathbf{t'},-z)={\rm Res}_z z^{M_2k-1}\psi_2(\mathbf{t},z)\psi_2(\mathbf{t'},-z),\quad k\geq1,\ odd.
  \end{align*}
  By the similar method in Proposition \ref{reductionbilinear}, we can know $\big(B_{M_1k}^{(1)}+B_{M_2k}^{(2)}\big)
  (\psi_a)=z^{M_ak}\psi_a$, that is $D_{M_1k,M_2k}(\psi_a)=z^{M_ak}\psi_a$ with $D_{M_1k,M_2k}=\pa_{t^{(1)}_{M_1k}}+\pa_{t^{(2)}_{M_2k}}$, which means $D_{M_1k,M_2k}(W_a)=0$ by $\psi_a(\bft,z)=W_a(\bft,\pa_a)\big(e^{\xi(t^{(a)},z)}\big)$. Thus the $(M_1,M_2)$-reduction of the 2-BKP hierarchy does not depend on $t^{(1)}_{M_1k}+t^{(2)}_{M_2k}$ $(k\geq1, odd)$.
\end{remark}
\begin{example}\label{ex:2n-22}
When $M_1=2n-2$, $M_2=2$, we have by (\ref{mathcallexpress}) that
\begin{align*}
  \mathcal{L}=\widetilde{B}_{2n-2}^{(1)}+\widetilde{v_0}+\pa_2^2.
\end{align*}
Recall from (\ref{BCjoint}), we know $\widetilde{B}_{2n-2}^{(1)*}=\pa_1B_{2n-2}^{(1)}\pa_1^{-1}$, thus if we denote 
$\widetilde{A}_{2n-1}^{(1)}=\pa_1B_{2n-2}^{(1)}$, then we can assume that
\begin{align*}
 \widetilde{A}_{2n-1}^{(1)}=\pa_1^{2n-1}+\sum_{l=1}^{n-1}(\widetilde{u}_l\pa_1^{2l-1}+\pa_1^{2l-1}\widetilde{u}_l),
\end{align*}
where $\widetilde{u}_1=-\sum_{l=2}^{n-1}\widetilde{u}_l^{(2l-2)}$. Therefore
\begin{align}\label{b2n-2}
 \widetilde{B}_{2n-2}^{(1)}=\pa_1^{2n-2}+\sum_{l=1}^{n-1
 }\pa_1^{-1}(\widetilde{u}_l\pa_1^{2l-1}+\pa_1^{2l-1}\widetilde{u}_l).
\end{align}
Further by Proposition \ref{2-bkplaxPi}, we have
 \begin{align}\label{pipa1pa22}
  \pi_1(\pa_2^2)=-\pa_1^{-1}\cdot\pa_2(\rho)+\pa_1^{-1}\rho\cdot\pa_1^{-1}\rho,
  \end{align}
and by $\pa_1(L_2^2)=-[\pa_2^{-1}\rho,L_2^2]$, we can find 
\begin{align}\label{pa1vo}
\pa_1(\widetilde{v_0})=2\pa_2(\rho).
\end{align}
So if we set $u_1=\widetilde{u}_1+\frac{v_0}{2}$, $u_i=\widetilde{u}_i$ $(i=2,3,\ldots,n-1)$, then by (\ref{b2n-2})-(\ref{pa1vo}), we finally have
\begin{align*}
    \pi_1(\mathcal{L})=\pa_1^{2n-2}+\sum_{l=1}^{n-1}\pa_1^{-1}(u_{l}\pa_1^{2l-1}+
    \pa_1^{2l-1}u_{l})+\pa_1^{-1}\rho\pa_1^{-1}\rho,
  \end{align*}
which is just the Lax operator of Drinfeld-Sokolov hierarchy of D type \cite{Drinfeld1984,Liu2011}.
\end{example}

\begin{example}
  When $M_1=3$, $M_2=1$, the corresponding Lax operator is $$\mathcal{L}=\pa_1^3+v_{1,0}\pa_1+\pa_2,$$ then by Proposition \ref{2-bkplaxPi}, we can know that
    $$\pi_1(\mathcal{L})=\pa_1^3+v_{1,0}\pa_1-\pa_1^{-1}\rho$$
 satisfies $\pi_1(\mathcal{L})^*=-\pa_1\cdot \pi_1(\mathcal{L})\cdot\pa_1^{-1}$.
\end{example}

\section{Transformations of the 2-BKP tau functions}\label{2bkpTransformations}
In this section, we firstly introduce the squared eigenfunction potential (SEP) for the 2-BKP hierarchy. Then based upon this, we discuss the transformations of the 2-BKP tau functions and the corresponding changes in the 2-BKP Lax triple. Finally we discuss the condition for the above transformations in the $(M_1,M_2)$-reduction.
\subsection{The squared eigenfunction potentials}\label{sepdef}
Firstly given the 2-BKP Lax triple $(L_1,L_2,H)$, the 2-BKP eigenfunction $q$ is defined by
\begin{align*}
\pa_{a,n}q=B_n^{(a)}(q),\quad H(q)=0,\quad a=1,2.
\end{align*}
Notice that $1$ is not the 2-BKP eigenfunction, which is different from the usual BKP case. 
\begin{lemma}\label{omegadef1}
 For two 2-BKP eigenfunctions $q_1$ and $q_2$,
 \begin{align}\label{omegadef}
 (q_2q_{1x_1}-q_1q_{2x_1})_{x_2}=-(q_2q_{1x_2}-q_1q_{2x_2})_{x_1}.
 \end{align} 
 \begin{proof}
   Firstly by direct computation,
   \begin{align*}
     (q_2q_{1x_1}-q_1q_{2x_1})_{x_2}+(q_2q_{1x_2}-q_1q_{2x_2})_{x_1}=2q_2\cdot\partial_1 \partial_2(q_{1})-2q_1\cdot\partial_1 \partial_2(q_{2}).
\end{align*}
Then by $H(q_{1})=H(q_{2})=0$, we can know $\partial_1 \partial_2(q_{1})=-\rho q_1$, $\partial_1 \partial_2(q_{2})=-\rho q_2$, leading to $q_2\cdot\partial_1 \partial_2(q_{1})-q_1\cdot\partial_1 \partial_2(q_{2})=0$.
 \end{proof}
\end{lemma}

Due to Lemma \ref{omegadef1}, we can define the SEP for the 2-BKP hierarchy. In fact, for the 2-BKP eigenfunction $q_1$ and $q_2$, the SEP $\Omega(q_1,q_2)$ is given by
\begin{equation}
\begin{aligned}\label{besp}
\Omega(q_1,q_2)&=\partial_1^{-1}(q_2q_{1x_1}-q_1q_{2x_1})=-\partial_2^{-1}(q_2q_{1x_2}-q_1q_{2x_2}).
\end{aligned}
\end{equation}
Notice that $\Omega(q_1,q_2)$ can be up to some constant. This SEP $\Omega(q_1,q_2)$ is the two dimensional generalization of the SEP in the usual BKP hierarchy\cite{Cheng2010,Loris1999}.
\begin{lemma}\label{omegaexpres}
  Given two 2-BKP eigenfunctions $q_1$ and $q_2$, 
  \begin{align*}
    &\partial_{a,n}\Omega(q_1,q_2)=C_n^{(a)}(q_1,\pa_a)(q_2), \quad a=1,2,
  \end{align*}
  where $C_n^{(a)}(q_1,\pa_a)=\sum_{j=0}^{n}C^{(a)}_{n,j}(q_1)\partial_a^j$.
  \begin{proof}
  Firstly, we can write $\Omega(q_1,q_2)$ into
  \begin{align}\label{omegaqua2}
   \Omega(q_1,q_2)=A_a[q_1](q_2),\quad a=1,2,
  \end{align}
  where $A_a[q_1]=(-1)^{a}\partial_a^{-1}\big(q_1\partial_a-q_{1x_a}\big)=(-1)^{a}\big(q_1-2\partial_a^{-1}q_{1x_a}\big)$.
   Then we have
   \begin{align*}
      \partial_{a,n}\Omega(q_1,q_2)&= 
      \big(\partial_{a,n}A_a[q_1]+A_a[q_1]\cdot B_n^{(a)}\big)(q_2).
    \end{align*}
   If we denote $C_n^{(a)}(q_1,\pa_a)=\partial_{a,n}A_a[q_1]+A_a[q_1]\cdot B_n^{(a)}$, then we can find
   \begin{align*}
      \big(C_n^{(a)}(q_1,\pa_a)\big)_{a,<0}=(-1)^{a-1}2\big(\partial_a^{-1}\partial_{a,n}(q_{1x_a})+\partial_a^{-1}q_{1x_a}B_n^{(a)}\big)_{a,<0}.
   \end{align*}
  By the formula $\big(\partial_a^{-1}\cdot f\cdot A(\partial_a)\big)_{<0}=\partial_a^{-1}\cdot A^*(f)+\partial^{-1}\cdot f\cdot A(\partial_a)_{<0}$, we have $\big(C_n^{(a)}(q_1,\pa_a)\big)_{a,<0}=0$ since $\partial_{a,n}(q_{1x_a})=-B_n^{(a)*}(q_{1x_a})$, where $B_n^{(a)*}=-\partial_aB_n^{(a)}\partial_a^{-1}$ is used. Further the highest order term of $C_n^{(a)}(q_1,\pa_a)$ lies in $(-1)^{a}q_1B_n^{(a)}$, which is $(-1)^{a}\pa_a^n$. So finally we can assume that $C_n^{(a)}(q_1,\pa_a)$ has the form 
 $C_n^{(a)}(q_1,\pa_a)=\sum_{j=0}^{n}C_{n,j}^{(a)}(q_1)\partial_a^j.$
  \end{proof}
\end{lemma}
\begin{lemma}\label{prospec}
  Given the 2-BKP eigenfunction $q(\mathbf{t})$ and the 2-BKP wave function $\psi_a(\mathbf{t},z)$, the SEP 
  $\Omega\big(q(\mathbf{t}),\psi_a(\mathbf{t},z)\big)$ has the following expansion in $z$,
  \begin{align*}
    &\Omega\big(q(\mathbf{t}),\psi_a(\mathbf{t},-z)\big)=(-1)^{a}\big(q(\mathbf{t})+\mathcal{O}(z^{-1})\big)e^{-\xi(t^{(a)},z)},\quad a=1,2,
    \end{align*}
    where the constant in the SEP $\Omega\big(q(\mathbf{t}),\psi_a(\mathbf{t},z)\big)$ involving $\psi_a(\mathbf{t},z)$ is fixed by the way below:
    \begin{align*}
      \Omega\big(q(\mathbf{t}),\psi_a(\mathbf{t}
      ,-z)\big)&=\big(A_a[q(\bft)]\cdot W_a\big)(e^{\xi(t^{(a)},z)}),
    \end{align*}
    and $\partial_a^{-k}(e^{\xi(t^{(a)},z)})=z^{-k}e^{\xi(t^{(a)},z)}$.
  \begin{proof}
    Firstly recall that $\psi_a(\mathbf{t},z)
=W_a(\mathbf{t},\partial_{a})(e^{\xi(t^{(a)},z)})$. So by (\ref{omegaqua2}), we can know
\begin{align*}
  \Omega\big(q(\mathbf{t}),\psi_a(\mathbf{t},-z)
  \big)&=\big(A_a[q(\bft)]\cdot W_a\big)(e^{\xi(t^{(a)},z)})
  =(-1)^{a}\big(q(\mathbf{t})+\mathcal{O}(z^{-1})\big)(e^{\xi(t^{(a)},z)}).
\end{align*}
  \end{proof}
\end{lemma}

\begin{proposition}\label{qtspectral}
  The 2-BKP eigenfunctions $q(\mathbf{t})$ has the following spectral representation,
  \begin{align}
& q(\mathbf{t})=\frac{1}{2}\sum_{a=1}^{2}{\rm Res}_z (-1)^a z^{-1}\psi_a(\mathbf{t},z)\Omega\big(
q(\mathbf{t'}),\psi_a(\mathbf{t'},-z)\big),\label{egienspe1} \end{align} 
        where $\psi_a(\mathbf{t},z)$ is the 2-BKP wave function.
        \begin{proof}
  Firstly, let us denote $$I(\mathbf{t},\mathbf{t'})=\frac{1}{2}\sum_{a=1}^{2} {\rm Res}_z (-1)^az^{-1}\psi_a(\mathbf{t},z)\Omega\big(
  q(\mathbf{t'}),\psi_a(\mathbf{t'},-z)\big).$$
  Then based on Lemma \ref{omegaexpres}, 
  \begin{align*}
   I(\mathbf{t},\mathbf{t'})_{t_n^{(b)'}}=\frac{1}{2}C_n^{(b)}(q_1(\bft'),\pa_{x'_b})\sum_{a=1}^{2}{\rm Res}_z (-1)^{a}z^{-1}\psi_a(\mathbf{t},z)\psi_a(\mathbf{t'},-z)=0,\quad b=1,2,
  \end{align*}
  where we have used the 2-BKP bilinear equation (\ref{bilineareq}). Therefore, $I(\mathbf{t},\mathbf{t'})$ is independent of $\mathbf{t'}$. So if set $\mathbf{t'}=\mathbf{t}$ in $I(\mathbf{t},\mathbf{t'})$ and use Lemma \ref{prospec}, we have 
    \begin{align*}
    I(\mathbf{t},\mathbf{t'})=I(\mathbf{t},\mathbf{t})=&\frac{1}{2}\sum_{a=1}^{2}{\rm Res}_z z^{-1}\psi_a(\mathbf{t},z)\big(q(\mathbf{t})+\mathcal{O}(z^{-1})\big)(e^{-\xi(t^{(a)},z)})
  =q(\mathbf{t}).
  \end{align*}
  \end{proof}
\end{proposition}

\begin{proposition}\label{omegaequa}
  Given the 2-BKP eigenfunction $q(\mathbf{t})$ and the 2-BKP wave function $\psi_a(\mathbf{t},z)$, the SEP 
  $\Omega\big(q(\mathbf{t}),\psi_a(\mathbf{t},z)\big)$ has the following expression,
  \begin{align}\label{eqigenrel}
  \Omega\big(q(\mathbf{t}),\psi_a(\mathbf{t},-z)\big)
=&(-1)^{a}\psi_a(\mathbf{t},-z)q(\mathbf{t}+2[z^{-1}]_a), \quad a=1,2.
  \end{align}
 And thus
  \begin{align}\label{egienspe2}
& q(\mathbf{t})= \frac{1}{2}\sum_{a=1}^{2}{\rm Res}_z z^{-1}\psi_a(\mathbf{t},z)\psi_a(\mathbf{t'},-z)q(\mathbf{t'}+2[z^{-1}]_a),\quad a=1,2. \end{align}
  \begin{proof}
  Firstly let us assume 
\begin{align}\label{expansion3}
  &\Omega\big(q(\mathbf{t}),\psi_a(\mathbf{t},-z)\big)=\widehat{h}_a(\mathbf{t},-z)e^{\xi(t^{(a)},-z)},
  \end{align}
then according to Lemma \ref{prospec}, $\widehat{h}_a(\mathbf{t},-z)$ has the form $(-1)^{a}\big(q(\mathbf{t})+\mathcal{O}(z^{-1})
\big)$.
If we set $\mathbf{t}\rightarrow\mathbf{t}+2[\lambda^{-1}]_1$, $\mathbf{t'}\rightarrow\mathbf{t}$ in (\ref{egienspe1}), we can obtain 
  \begin{align*}
     \widehat{h}_1(\mathbf{t},-\lambda)=-q(\mathbf{t}+2[\lambda^{-1}]_1)\frac{\tau(\mathbf{t}+2[\lambda^{-1}]_1)}{\tau(\mathbf{t})},
   \end{align*}
  where (\ref{psitaurelation}) is used and $\tau(\mathbf{t})$ is the 2-BKP tau function. So we have
  \begin{align*}
     \Omega\big(q(\mathbf{t}),\psi_1(\mathbf{t},-z)\big)&=-\psi_1(\mathbf{t},-z)q(\mathbf{t}+2[z^{-1}]_1).
   \end{align*}
     Similarly $\Omega\big(q(\mathbf{t}),\psi_2(\mathbf{t},-z)\big)$ can be obtained by setting $\mathbf{t}\rightarrow\mathbf{t}+2[\lambda^{-1}]_2$, $\mathbf{t'}\rightarrow\mathbf{t}$ in (\ref{egienspe1}).
  \end{proof}
\end{proposition}
\begin{remark}
  Given the 2-BKP eigenfunction $q(\mathbf{t})$ and the 2-BKP tau function 
  $\tau(\mathbf{t})$, we can find by (\ref{expansion3}) that $\big(\tau_0(\mathbf{t}),\tau_1(\mathbf{t})\big)=\big(\tau(\mathbf{t}),q(\mathbf{t})\tau(\mathbf{t})\big)$
satisfies the 2-modified BKP (2-mBKP) hierarchy \cite{Jimbo1983,van2014}
\begin{equation*}
    \begin{aligned}
        &{\rm Res}_z z^{-1}\tau_0(\mathbf{t}-2[z^{-1}]_1)\tau_1(\mathbf{t}'+2[z^{-1}]_1)e^{\xi(t^{(1)}-t^{(1)'},z)}\\&+{\rm Res}_z z^{-1}\tau_0(\mathbf{t}-2[z^{-1}]_2)\tau_1(\mathbf{t}'+2[z^{-1}]_2)e^{\xi(t^{(2)}-t^{(2)'},z)}
=2\tau_1(\mathbf{t})\tau_0(\mathbf{t}').
    \end{aligned}
    \end{equation*}
    Further by ${\rm Res}_z a(-z)=-{\rm Res}_z a(z)$, we can know $\big(\tau_0(\mathbf{t}),\tau_1(\mathbf{t})\big)=\big(q(\mathbf{t})\tau(\mathbf{t}),\tau(\mathbf{t})\big)$
 also satisfies the 2-mBKP hierarchy.
 \end{remark}
\begin{corollary}\label{qtpsit}
  Given the 2-BKP eigenfunction $q(\mathbf{t})$ and the 2-BKP wave function $\psi_a(\mathbf{t},z)$ ($a=1,2$), we have
  \begin{itemize}\label{eigen2}
\item {$\big(q(\mathbf{t}-2[z^{-1}]_a)\psi_a(\mathbf{t},z)\big)_{x_a}=q^2(\mathbf{t})
    (q(\mathbf{t})^{-1}\psi_a(\mathbf{t},z))_{x_a}$},
\item {$\big(q(\mathbf{t}-2[z^{-1}]_a)\psi_{a}(\mathbf{t},z)\big)_{x_{3-a}}=-q^2(\mathbf{t})
    \big(q(\mathbf{t})^{-1}\psi_a(\mathbf{t},z)\big)_{x_{3-a}}$}.
\end{itemize}
\begin{proof}
  This corollary can be proved directly by (\ref{besp}) and (\ref{eqigenrel}).
\end{proof}
\end{corollary}
\begin{lemma}\label{lemmaomega}
  Given the 2-BKP eigenfunctions $q_1$ and $q_2$,
  \begin{align*}
   \frac{1}{2}\sum_{a=1}^{2}(-1)^{a}{\rm Res}_z z^{-1}\Omega\big(q_1(\mathbf{t}),\psi_a(\mathbf{t},z)\big)\Omega\big(q_2(\mathbf{t'}),\psi_a(\mathbf{t'},-z)\big)
       =\Omega\big(q_1(\mathbf{t}),q_2(\mathbf{t})\big)-\Omega\big(q_1(\mathbf{t'}),q_2(\mathbf{t'})\big).
         \end{align*}
         \begin{proof}
           If we denote $F(\mathbf{t},\mathbf{t'})=\frac{1}{2}\sum_{a=1}^{2}(-1)^{a}{\rm Res}_z z^{-1}\Omega\big(q_1(\mathbf{t}),\psi_a(\mathbf{t},z)\big)\Omega\big(q_2(\mathbf{t'}),\psi_a(\mathbf{t'},-z)\big)
    $, then we obtain
    \begin{align*}
      \partial_{t_m^{b}}\partial'_{t_n^{c}}F(\mathbf{t},\mathbf{t'})&=\frac{1}{2}
      C_m^{(b)}(q_1(\bft),\pa_{b})C_n^{(c)}(q_2(\bft'),\pa'_{c})
      \sum_{a=1}^{2}(-1)^{a}{\rm Res}_z z^{-1}\psi_a(\bft,z)\psi_a(\bft',-z)=0,   \end{align*}
    where $\pa'_{c}=\pa_{x'_c}$ and we have used Lemma \ref{omegaexpres} and (\ref{bilineareq}). Thus we find $F(\mathbf{t},\mathbf{t'})$ has the form of $f(\mathbf{t})+g(\mathbf{t'})$ for some $f$ and $g$ to be determined. Also, if setting $\bft'=\bft$, then by Lemma \ref{prospec}, we have $F(\bft,\bft)=0$. Thus, we have $f(\bft)=-g(\bft)$. If we denote $\mathbf{r}=\bft-\bf{p}$, $\mathbf{r'}=\bft+\mathbf{p}$, then we have
     \begin{align*}
     2g(\bft)_{t_n^{(b)}}&=\pa_{p_n^{(b)}}F(\mathbf{r},\mathbf{r'})|_{\mathbf{p}=0}
     =-C_{n}^{b}(q_1(\mathbf{t}),\pa_b)\big(q_2(\mathbf{t})\big)+C^{b}_{n}(q_2(\mathbf{t}),\pa_b)\big(q_1(\mathbf{t})\big)=-2\pa_{t_n^{(b)}}
     \Omega\big(q_1(\bft),q_2(\bft)\big), \end{align*}
    where we used Lemma \ref{omegaexpres}. Thus finally we derive $$F(\mathbf{t},\mathbf{t'})=\Omega\big(q_1(\bft),q_2(\bft)\big)-\Omega\big(q_1(\bft'),q_2(\bft')\big).$$
         \end{proof}
\end{lemma}
\begin{proposition}\label{omega12}
 Given the 2-BKP eigenfunctions $q_1$ and $q_2$, if denote $\Omega_{12}(\bft)=\Omega\big(q_1(\bft),q_2(\bft)\big)$, then
 \begin{align*}
 \Omega_{12}(\mathbf{t}-2[\lambda^{-1}]_a)-\Omega_{12}(\bft)=(-1)^{a-1}
    \Big(q_2(\mathbf{t})q_1(\mathbf{t}-2[\lambda^{-1}]_a)-q_1(\mathbf{t})q_2(\mathbf{t}-2[\lambda^{-1}]_a)\Big),\quad a=1,2. \end{align*}
         \begin{proof}
           Based on Lemma \ref{lemmaomega}, we can prove this proposition by setting $\mathbf{t}-\mathbf{t'}=2[\lambda^{-1}]_a$.
         \end{proof}
\end{proposition}
\subsection{Proof of Theorem \ref{mainqtau}}\label{proofth1}
After the preparation above, now we can prove {\bf{Theorem \ref{mainqtau}}}. Firstly starting from (\ref{egienspe2}), we can know
\begin{align}\label{prooftheorem1}
  \sum_{a=1}^{2}{\rm Res}_zz^{-1} q(\mathbf{t})^{-1}\psi_a(\mathbf{t},z)\psi_a(\mathbf{t'},-z)q(\mathbf{t'}+2[z^{-1}]_a)=2.
\end{align}
If apply $\partial_{x_1}$ to (\ref{prooftheorem1}) and use Corollary \ref{qtpsit}, we have
\begin{align*}
  \sum_{a=1}^{2}(-1)^{a-1}{\rm Res}_z z^{-1} \big(q(\mathbf{t}-2[z^{-1}]_a)\psi_a(\mathbf{t},z)\big)_{x_1}\psi_a(\mathbf{t'},-z)q(\mathbf{t'}+2[z^{-1}]_a)=0.
\end{align*}
Thus if set $\psi^{[1]}_a(\mathbf{t},z)=\frac{\tau^{[1]}(\mathbf{t}-2[z^{-1}]_a)}{\tau^{[1]}(\mathbf{t})}e^{\xi(t^{(a)},z)}$ with $\tau^{[1]}(\mathbf{t})=q(\mathbf{t})\tau(\mathbf{t})$, then 
\begin{align}\label{prooftheorem3}
  \sum_{a=1}^{2}(-1)^{a-1}{\rm Res}_z z^{-1} \big(\psi^{[1]}_a(\mathbf{t},z)q(\mathbf{t})\big)_{x_1}\psi^{[1]}_a(\mathbf{t'},-z)=0.
\end{align}

Let us define $W^{[1]}_a=1+\sum_{i=1}^{+\infty}w^{[1]}_{a,i}\partial_a^{-i}$ such that 
\begin{align*}
  \psi^{[1]}_a(\mathbf{t},z)=W^{[1]}_a(\mathbf{t},\partial_a)\big(e^{\xi(t^{(a)},z)}\big),\quad a=1,2,
\end{align*}
then if set $\mathbf{t'}=\mathbf{t}$ in (\ref{prooftheorem3}) except $x'_1$, then we have
\begin{align*}
{\rm Res}_z \big(\partial_{1} q(\mathbf{t})W^{[1]}_1(\mathbf{t},\partial_1)\partial_{1}^{-1}\big)(e^{x_1z})\cdot W^{[1]}_1(\mathbf{t'},\partial_1)(e^{-x'_1z})=q(\mathbf{t})_{x_1}.
\end{align*}
So by Lemma \ref{lemma:ref5lemma4}, we have
\begin{align*}
  \big(\partial_{1} q(\mathbf{t})W^{[1]}_1(\mathbf{t},\partial_1)\partial_{1}^{-1}W^{[1]}_1(\mathbf{t},\partial_1)^*\partial_{1}\big)\big((x_1-x'_1)^{0}\big)= q(\mathbf{t})_{x_1},
\end{align*}
implying by Remark \ref{remarkaxbx} that
$$ \big(\partial_{1}q(\mathbf{t})W^{[1]}_1(\mathbf{t},\partial_1)\partial_{1}^{-1}W^{[1]}_1(\mathbf{t},\partial_1)^*\partial_{1}\big)_{1,\leq0}=q(\mathbf{t})_{x_1}.$$
Thus $\partial_{1} q(\mathbf{t})\cdot W^{[1]}_1(\mathbf{t},\partial_1)\partial_{1}^{-1}W^{[1]}_1(\mathbf{t},\partial_1)^*\partial_{1}=q(\mathbf{t})_{x_1}
+q(\mathbf{t})\partial_1=\partial_1q(\mathbf{t})$, and we have
\begin{align}\label{prooftheorem4}
  W^{[1]}_1(\mathbf{t},\partial_1)\partial_{1}^{-1}W^{[1]}_1(\mathbf{t},\partial_1)^*=\partial_1^{-1}.
\end{align}

If apply $\partial_{1,n}$ to (\ref{prooftheorem3}) and set $\mathbf{t'}=\mathbf{t}$ in (\ref{prooftheorem3}) except $x'_1$, similarly we have 
\begin{align*}
  \partial_{1,n}q_{x_1}&=\big(\partial_{1,n}(\partial_1qW^{[1]}_1)+\partial_1qW^{[1]}_1\partial_1^n\big)(\partial_1^{-1}W^{[1]*}_1\partial_1)\big((x_1-x'_1)^{0}\big).
  \end{align*}
Next by (\ref{prooftheorem4}), we have
\begin{equation}\label{prooftheorem5}
  \begin{aligned}
  \partial_{1,n}q_{x_1}&=\big(\partial_{1,n}(\partial_1qW^{[1]}_1)\cdot W^{[1]-1}_1\big)_{1,\leq0}+
  \big(\partial_1qW^{[1]}_1\partial_1^nW^{[1]-1}_1\big)_{1,\leq0}\\&
  =\partial_{1,n}(\partial_1qW^{[1]}_1)\cdot W^{[1]-1}_1-\big(\partial_{1,n}(\partial_1qW^{[1]}_1)\cdot W^{[1]-1}_1\big)_{1,>0}+  \big(\partial_1qW^{[1]}_1\partial_1^nW^{[1]-1}_1\big)_{1,\leq0}
\\&=\partial_{1}\cdot\partial_{1,n}q+\partial_{1}q\cdot \partial_{1,n}W^{[1]}_1\cdot W^{[1]-1}_1-\partial_{1,n}q\cdot\partial_{1}+\big(\partial_1qW^{[1]}_1\partial_1^nW^{[1]-1}_1\big)_{1,\leq0},
\\&=\partial_{1,n}q_{x_1}+\partial_{1}q\cdot \partial_{1,n}W^{[1]}_1\cdot W^{[1]-1}_1+\big(\partial_1qW^{[1]}_1\partial_1^nW^{[1]-1}_1\big)_{1,\leq0},
\end{aligned}
\end{equation}
where in the third equality, one should note that $\big(\partial_{1,n}(\partial_1qW^{[1]}_1)\cdot W^{[1]-1}_1\big)_{1,>0}=\partial_{1,n}(q)\partial_1$. Now
(\ref{prooftheorem5}) will become
\begin{align*}
  \partial_{1}q\cdot \partial_{1,n}W^{[1]}_1\cdot W^{[1]-1}_1&=-\big(\partial_1\cdot q\cdot W^{[1]}_1\partial_1^nW^{[1]-1}_1\big)_{1,\leq0}
  =-\big(\partial_1q\cdot (W^{[1]}_1\partial_1^nW^{[1]-1}_1)_{1,\leq0}\big)_{1,\leq0}\\&
  =-\big(\partial_1 q\cdot(W^{[1]}_1\partial_1^nW^{[1]-1}_1)_{1,<0}\big)_{1,\leq0}
  =-\partial_1q\cdot(W^{[1]}_1\partial_1^nW^{[1]-1}_1)_{1,<0}.
\end{align*}
Here we used the fact that $(W^{[1]}_1\partial_1^nW^{[1]-1}_1)_{1,[0]}=0$ if $W^{[1]}_1\partial_1^{-1}W^{[1]*}_1=\partial_1^{-1}$. So finally we can obtain 
$$\partial_{1,n}W^{[1]}_1=-(W^{[1]}_1\partial_1^nW^{[1]-1}_1)_{1,<0}W^{[1]}_1.$$

If we apply $\partial_{x_2}$ to (\ref{prooftheorem1}) and use Corollary \ref{qtpsit}, we can obtain 
\begin{align}\label{profftheorem6}
   \sum_{a=1}^{2}(-1)^{a}{\rm Res}_z z^{-1} \big(\psi^{[1]}_a(\mathbf{t},z)q(\mathbf{t})\big)_{x_2}\psi^{[1]}_a(\mathbf{t'},-z)=0.
\end{align}
 Similar discussions can lead to 
 \begin{align*}
  &W^{[1]}_2\partial_{2}^{-1}W^{[1]*}_2=\partial_2^{-1},\quad \partial_{2,n}W^{[1]}_2=-(W^{[1]}_2\partial_2^nW^{[1]-1}_2)_{2,<0}W^{[1]}_2.
\end{align*}

 Now if applying $\partial_{2,n}$ to (\ref{prooftheorem3}) and letting $\mathbf{t'}=\mathbf{t}$ except $x'_1$, we can obtain by Lemma \ref{lemma:ref5lemma4} and $W^{[1]}_a\partial_a^{-1}W^{[1]*}_a=\partial_a^{-1}$ that
 \begin{align*}
   &\bigg(\partial_{2,n}\big(\partial_1\cdot q(\mathbf{t})\cdot W^{[1]}_1(\mathbf{t},\partial_1)\big)W^{[1]}_1(\mathbf{t},\partial_1)^{-1}\bigg)\big((x_1-x'_1)^{0}\big)
   \\&=\bigg(\big(\partial_{2,n}\big(q(\mathbf{t})W^{[1]}_2(\mathbf{t},\partial_2)\big)_{x_1}+\big(q(\mathbf{t})W^{[1]}_2(\mathbf{t},\partial_2)\big)_{x_1}\cdot\partial_2^n\big)W^{[1]}_2(\mathbf{t'},\partial_2)^{-1}\bigg)_{2,[0]}
   \\&=\partial_{2,n}q(\mathbf{t})_{x_1}+\bigg(\big(q(\mathbf{t})W^{[1]}_2(\mathbf{t},\partial_2)\big)_{x_1}\cdot\partial_2^nW^{[1]}_2(\mathbf{t'},\partial_2)^{-1}\bigg)_{2,[0]}.
     \end{align*}
     So by Remark \ref{remarkaxbx} and Lemma \ref{dressinglemma},
     \begin{align}\label{proofth11}
      \big(\partial_{2,n}(\partial_1\cdot q\cdot W^{[1]}_1)W^{[1]-1}_1\big)_{1,\leq0}=\partial_{2,n}q_{x_1}+\big((qW^{[1]}_2)_{x_1}\cdot\partial_2^n\partial_1^{-1}W^{[1]-1}_2\partial_1\big)_{2,[0]}.
     \end{align}
     Notice that
      \begin{align*}
   &\bigg(\partial_{2,n}\big(\partial_1\cdot q\cdot W^{[1]}_1\big)\cdot W^{[1]-1}_1\bigg)_{1,\leq0}
   =\partial_{2,n}\big(\partial_1\cdot q\cdot W^{[1]}_1\big)\cdot W^{[1]-1}_1-\partial_{2,n}q\partial_1
   \\&=\partial_1\cdot q\cdot \partial_{2,n}W^{[1]}_1\cdot W^{[1]-1}_1+\partial_1\cdot\partial_{2,n}q-\partial_{2,n}q\partial_1\\&=
   \partial_1\cdot q\cdot \partial_{2,n}W^{[1]}_1\cdot W^{[1]-1}_1+\partial_{2,n}q_{x_1}.
   \end{align*}
 So by (\ref{proofth11}) and $(qW^{[1]}_2)_{x_1}=\partial_{1}qW^{[1]}_2-qW^{[1]}_2\partial_{1}$, we can know  
 \begin{align*}
\partial_1q\cdot \partial_{2,n}W^{[1]}_1\cdot W^{[1]-1}_1&=\big((qW^{[1]}_2)_{x_1}\cdot\partial_2^n\partial_1^{-1}W^{[1]-1}_2\partial_1\big)_{2,[0]}\\
   &=\partial_1q\cdot \big(W^{[1]}_2\partial_2^n\partial_1^{-1}W^{[1]-1}_2\partial_1\big)_{2,[0]}
   -q\big(W^{[1]}_2\partial_2^nW^{[1]-1}_2\big)_{2,[0]}\partial_1\\
   &=\partial_1q\cdot   \big(W^{[1]}_2\partial_2^n\partial_1^{-1}W^{[1]-1}_2\partial_1\big)_{2,[0]},
 \end{align*}
 where $\big(W^{[1]}_2\partial_2^nW^{[1]-1}_2\big)_{2,[0]}=0$ if $W^{[1]}_2\partial_{2}^{-1}W^{[1]*}_2=\partial_2^{-1}$. And thus 
 \begin{align*}
&\partial_{2,n}W^{[1]}_{1}=(W^{[1]}_2\partial_2^n\partial_{1}^{-1}W^{[1]-1}_2\partial_{1})_{2,[0]}W^{[1]}_{1}.
\end{align*}
 Similarly, if applying $\partial_{1,n}$ to (\ref{profftheorem6}) and letting $\mathbf{t'}=\mathbf{t}$ except $x'_2$, we can derive that
\begin{align*}
&\partial_{1,n}W^{[1]}_{2}=(W^{[1]}_1\partial_1^n\partial_{2}^{-1}W^{[1]-1}_1\partial_{2})_{1,[0]}W^{[1]}_{2}.
\end{align*}

By now, we have obtained the following relations for $W^{[1]}_a$ ($a=1,2$):
\begin{equation}\label{wsatitdsa}
\begin{aligned}
&W^{[1]}_a\partial_a^{-1}W^{[1]*}_a\partial_a=1,\\ 
&\partial_{a,n}W^{[1]}_{a}=-(W^{[1]}_a\partial_a^nW^{[1]-1}_a)_{a,<0}W^{[1]}_a,\\ &\partial_{a,n}W^{[1]}_{3-a}=(W^{[1]}_a\partial_a^n\partial_{3-a}^{-1}W^{[1]-1}_a\partial_{3-a})_{a,[0]}W^{[1]}_{3-a}.
\end{aligned}\end{equation}
Further by Corollary \ref{Hexist}, we know that there exists $H^{[1]}=\pa_1\pa_2+\rho^{[1]}$ such that 
\begin{align}\label{H[1]}
H^{[1]}=\partial_1W^{[1]}_1\partial_{2}W^{[1]-1}_1=\partial_2W^{[1]}_2\partial_{1}W^{[1]-1}_2,
\end{align}
with $\rho^{[1]}=2\partial_1\partial_2(\log\tau^{[1]})$.
    And by Proposition \ref{relationdress},
\begin{align*}
  \partial_{a,n}W_{3-a}=(W_a\partial_a^nW_a^{-1})_{a,\geq0}(W_{3-a}).
\end{align*}
So by Theorem 3.9 in \cite{Geng2023}, we can know that 
\begin{align*}
\psi^{[1]}_a(\mathbf{t},z)&=W_{a}^{[1]}(\mathbf{t},\partial_a)\big(e^{\xi(t^{(a)},z)}\big)
=\frac{\tau^{[1]}(\mathbf{t}-2[z^{-1}]_a)}{\tau^{[1]}(\mathbf{t})}e^{\xi(t^{(a)},z)}
\end{align*}
 satisfies the 2-BKP bilinear equation,
  \begin{align*}
     {\rm Res}_z z^{-1}\psi_1^{[1]}(\mathbf{t},z)\psi_1^{[1]}(\mathbf{t'},-z)={\rm Res}_z z^{-1}\psi_2^{[1]}(\mathbf{t},z)\psi_2^{[1]}(\mathbf{t'},-z),
  \end{align*}   
  which means 
  \begin{equation*}
    \begin{aligned}
        &{\rm Res}_z z^{-1}\tau^{[1]}(\mathbf{t}-2[z^{-1}]_1)\tau^{[1]}(\mathbf{t}'+2[z^{-1}]_1)e^{\xi(t^{(1)}-t^{(1)'},z)}
\\&={\rm Res}_z z^{-1}\tau^{[1]}(\mathbf{t}-2[z^{-1}]_2)\tau^{[1]}(\mathbf{t}'+2[z^{-1}]_2)e^{\xi(t^{(2)}-t^{(2)'},z)}.
    \end{aligned}
\end{equation*}
  Therefore $\tau^{[1]}=q\tau$ is the 2-BKP tau function. Here we have finished the first part of Theorem \ref{mainqtau}.

For the second part of Theorem \ref{mainqtau}, let us see the lemma firstly.
\begin{lemma}\label{psi[1]}
  Given the 2-BKP tau function $\tau(\mathbf{t})$ and the 2-BKP eigenfunction $q(\bft)$, if denote $\tau^{[1]}(\bft)=q(\bft)\tau(\bft)$ and $\psi_a^{[1]}(\bft,z)=\frac{\tau^{[1]}(\mathbf{t}-2[z^{-1}]_a)}{\tau^{[1]}(\mathbf{t})}e^{\xi(t^{(a)},z)}$, then 
  \begin{align*}
    \psi_a^{[1]}(\bft,z)=(-1)^{a}\frac{\Omega\big(q(\mathbf{t}),\psi_a(\mathbf{t},z)\big)}{q(\mathbf{t})},
  \end{align*}
  where $\psi_a(\bft,z)$ is the 2-BKP wave function related with $\tau(\bft)$ by (\ref{wavefunction}).
  \begin{proof}
    Firstly it is obviously that
    \begin{align*}
  \psi^{[1]}_a(\mathbf{t},z)&=\frac{q(\mathbf{t}-2[z^{-1}]_a)\tau(\mathbf{t}-2[z^{-1}]_a)}{q(\mathbf{t})\tau(\mathbf{t})}e^{\xi(t^{(a)},z)}
 =\frac{q(\mathbf{t}-2[z^{-1}]_a)\psi_a(\mathbf{t},z)}{q(\mathbf{t})}.
\end{align*}
Then by (\ref{eqigenrel}), we can prove this lemma.
  \end{proof}
\end{lemma}
Based upon this lemma, if one denotes 
\begin{align*}
  T_b[q]=(-1)^{b}q^{-1}A_b[q]=1-2q^{-1}\partial_b^{-1}q_{x_b},
\end{align*}
where $A_b[q]$ is given in the proof of Lemma \ref{omegaexpres}, then by (\ref{omegaqua2}), we have
\begin{align*}
  \psi^{[1]}_a(\mathbf{t},z)=T_a[q]\big(\psi_a(\mathbf{t},z)\big),\quad a=1,2.
\end{align*}
Recall that $\psi^{[1]}_a(\mathbf{t},z)=W^{[1]}_a(\mathbf{t},\partial_a)\big(e^{\xi(t^{(a)},z)}\big)$, then by the fact that if $A(\bft,\pa_a)\big(e^{\xi(t^{(a)},z)}\big)=0$ for $A(\bft,\pa_a)\in \E_a^{0}$, then $A(\bft,\pa_a)=0$, we can know
\begin{align}\label{darbouxwave}
  W^{[1]}_a(\mathbf{t},\pa_a)=T_a[q]W_a(\mathbf{t},\pa_a),\quad a=1,2.
\end{align}
If denote the Lax operator $L_a^{[1]}=W_a^{[1]}\pa_aW_a^{[1]-1}$, then by Proposition \ref{Laxeqderive}, we have 
\begin{align*}
  \partial_{a,n}L_b^{[1]}=[\pi_{b}(B_n^{(a)[1]}),L_{b}^{[1]}], \quad a,b=1,2.
\end{align*}
where $B_n^{(a)[1]}=(L_{a}^{[1]})^n_{a,\geq0}$. Here $L_{a}^{[1]}$ is related with $L_{a}$ by
\begin{align*}
  L^{[1]}_a=T_a[q]L_aT_a[q]^{-1}, \quad a=1,2.
\end{align*}

Also by (\ref{wsatitdsa}) and (\ref{H[1]}), we have
\begin{align*}
 \partial_{a,n}H^{[1]}=-(B^{(a)[1]}_{n})^*H^{[1]}-H^{[1]}B^{(a)[1]}_{n}.
\end{align*}
And $H^{[1]}$ is connected with $H$ by 
\begin{align}\label{h[1]11}
  H^{[1]}=\pa_1T_1[q]\pa_1^{-1}HT_1[q]^{-1}
  =\pa_2T_2[q]\pa_2^{-1}HT_2[q]^{-1},
\end{align}
where we have substituted (\ref{darbouxwave}) into (\ref{H[1]}). Thus we have finished the proof of {\bf{Theorem \ref{mainqtau}}}.

If $\widetilde{q}$ is another 2-BKP eigenfunction and denote $\widetilde{q}^{[1]}=q^{-1}\Omega(\widetilde{q},q)$, then by (\ref{besp}), 
\begin{align*}                                                                                                                                                                                                                                                                                                                                                                                                                                                                                                                                                                                                                                                                                                                                                                                                                                                                  
  \widetilde{q}^{[1]}&=T_1[q](\widetilde{q})=-T_2[q](\widetilde{q}).
\end{align*}
By (\ref{darbouxwave}) and $\pa_{a,n}W_a=B_n^{(a)}W_a-W_a\pa_a^n$, we can know
\begin{align}\label{Bna[1]}
B_n^{(a)[1]}=T_a[q]B_n^{(a)}T_a[q]^{-1}+\pa_{a,n}T_a[q]\cdot T_a[q]^{-1}.
\end{align}
Therefore from (\ref{Bna[1]}), we can know
\begin{align*}
 \pa_{a,n}\widetilde{q}^{[1]}=B_n^{(a)[1]}(\widetilde{q}^{[1]}).
\end{align*}
Next by (\ref{h[1]11}) and $H(\widetilde{q})=0$, we can find $H^{[1]}(\widetilde{q}^{[1]})=0$. Therefore $\widetilde{q}^{[1]}$ is the 2-BKP eigenfunction corresponding to the Lax triple $(L_1^{[1]},L_2^{[1]},H^{[1]})$. Further we can know $\widetilde{q}^{[1]}\tau^{[1]}=\Omega(\widetilde{q},q)\tau$ is another 2-BKP tau function by Theorem \ref{mainqtau}. So we have proved {\bf{Corollary \ref{qtauistau}}}.
\begin{remark}\label{remarkpsi2diff}
  When $\widetilde{q}=\psi_2(\bft,z)$, we can find that there is a sign difference between our definition $\widetilde{q}=q^{-1}\Omega(\widetilde{q},q)$ and the one in Lemma \ref{psi[1]}. But the 2-BKP eigenfunction can be up to the multiplication of some constant. Therefore the definition of $\widetilde{q}=q^{-1}\Omega(\widetilde{q},q)$ is reasonable. In what follows, when discussing the 2-BKP wave functions $\psi_a(\bft,z)$, we will consider the corresponding signs, that is using Lemma \ref{psi[1]}, while for general 2-BKP eigenfunctions, we will ignore the signs.
\end{remark}
\subsection{Transformations in the reduction case}

In this subsection, we will try to prove Proposition \ref{m1m2redction} and Corollary \ref{mnreduction1}, which are just the reduction case. For this we need the lemma below.
\begin{lemma}\label{Omega12chanes}For the $(M_1,M_2)$-reduction of the 2-BKP hierarchy,
  given the tau function $\tau(\mathbf{t})$ and the eigenfunction $q(\mathbf{t})$, we have
  \begin{equation}\label{omega12exchange}
     \begin{aligned}
      &\Omega\Big(q_1(\mathbf{t}),\mathcal{L}\big(q_2(\mathbf{t})\big)\Big)
      +(-1)^{M_a}\Omega\Big(q_2(\mathbf{t'}),\mathcal{L}\big(q_1(\mathbf{t'})\big)\Big)
      \\&=\frac{1}{2}\sum_{a=1}^{2}(-1)^a{\rm Res}_z z^{M_a-1}\Omega\big(q_1(\mathbf{t}),\psi_a(\mathbf{t},z)\big)\Omega\big(q_2(\mathbf{t'}),\psi_a(\mathbf{t'},-z)\big),
     \end{aligned}
  \end{equation}
  where $(-1)^{M_1}=(-1)^{M_2}$ since $M_1+M_2$ is even.
 \begin{proof}
    Firstly, applying $\mathcal{L}$ to (\ref{egienspe1}) and using $\mathcal{L}(\psi_a)=z^{M_a}\psi_a$, we have
    \begin{align*}
      \mathcal{L}\big(q(\mathbf{t})\big)=\frac{1}{2}\sum_{a=1}^{2}(-1)^a{\rm Res}_z z^{M_a-1}\psi_a(\mathbf{t},z)\Omega\big(q(\mathbf{t'}),\psi_a(\mathbf{t'},-z)\big).
    \end{align*}
Then by (\ref{omegaqua2}), we can know
\begin{equation}\label{lpsiqrealtion1}
  \begin{aligned}
  &\Omega\Big(q_1(\mathbf{t}),\mathcal{L}\big(q_2(\mathbf{t})\big)\Big)=A_b[q_1(\bft)]
  \mathcal{L}\big(q_2(\mathbf{t})\big)\\&=
      \frac{1}{2}\sum_{a=1}^{2}(-1)^a{\rm Res}_z z^{M_a-1}A_b[q_1(\bft)]\mathcal{L}\big(\psi_a(\mathbf{t},z)\big)
      \Omega\big(q_2(\mathbf{t'}),\psi_a(\mathbf{t'},-z)\big)-c_{12}(\bft').
\end{aligned}
\end{equation}
Recall that $A_b[q(\bft)]=(-1)^{b}\partial_b^{-1}\big(q\partial_b-q_{x_b}\big)$ and $c_{12}(\bft')$ is the integrel constant, implying \[J(\bft,\bft')=\Omega\Big(q_1(\mathbf{t}),\mathcal{L}\big(q_2(\mathbf{t})\big)\Big)
+c_{12}(\bft'),\] 
where $J(\bft,\bft')$ is the right hand side of (\ref{omega12exchange}).

If set $\bft\leftrightarrow\bft'$, $q_1\leftrightarrow q_2$, $z\leftrightarrow -z$ in (\ref{lpsiqrealtion1}),
\begin{align}
  J(\bft,\bft')&=(-1)^{M_a}\bigg(\Omega\Big(q_2(\mathbf{t'}),\mathcal{L}\big(q_1(\mathbf{t'})
  \big)\Big)+c_{21}(\bft)\bigg),\label{lpsiqrealtion2}
\end{align}
where we should note that $(-1)^{M_1}=(-1)^{M_2}$ since $M_1+M_2$ is even, and the constant $c_{21}$ is different from $c_{12}$. Thus from (\ref{lpsiqrealtion1}) and (\ref{lpsiqrealtion2}), we have
    \begin{align*}
     \Omega\Big(q_1(\mathbf{t}),\mathcal{L}\big(q_2(\mathbf{t})\big)\Big)+c_{12}(\bft')
     =(-1)^{M_a}\bigg(\Omega\Big(q_2(\mathbf{t'}),\mathcal{L}\big(q_1(\mathbf{t'})\big)\Big)+c_{21}(\bft)\bigg),
    \end{align*}
which means
\begin{align*}
      \Omega\Big(q_1(\mathbf{t}),\mathcal{L}\big(q_2(\mathbf{t})\big)\Big)
      -(-1)^{M_a}c_{21}(\bft)=(-1)^{M_a}\Omega\Big(q_2(\mathbf{t'}),\mathcal{L}
      \big(q_1(\mathbf{t'})\big)\Big)-c_{12}(\bft')
    \end{align*}
is some constant. If denote this constant to be $c$, then 
\begin{align*}
  &c_{12}(\bft')=(-1)^{M_a}\Omega\Big(q_2(\mathbf{t'}),\mathcal{L}\big(q_1(\mathbf{t'})\big)\Big)-c,
  \quad c_{21}(\bft)=(-1)^{M_a}\bigg(\Omega\Big(q_1(\mathbf{t}),\mathcal{L}
  \big(q_2(\mathbf{t})\big)\Big)-c\bigg).
\end{align*}
So finally if set $\Omega\Big(q_1(\mathbf{t}),\mathcal{L}\big(q_2(\mathbf{t})\big)\Big)-c$ to be a new $\Omega\Big(q_1(\mathbf{t}),\mathcal{L}\big(q_2(\mathbf{t})\big)\Big)$, therefore we can get (\ref{omega12exchange}).
 \end{proof}
\end{lemma}

For the 2-BKP tau function $\tau(\bft)$ and the 2-BKP eigenfunction $q(\bft)$, if $\mathcal{L}\big(q(\bft)\big)=cq(\bft)$ for the $(M_1,M_2)$-reduction of the 2-BKP hierarchy, then by Lemma \ref{Omega12chanes}, we can know
\begin{align*}
\frac{1}{2}\sum_{a=1}^{2}(-1)^a{\rm Res}_z z^{M_a-1}\Omega\big(q(\mathbf{t}),\psi_a(\mathbf{t},z)\big)\Omega
\big(q(\mathbf{t'}),\psi_a(\mathbf{t'},-z)\big)=0.
\end{align*}
Further by (\ref{eqigenrel}), we can find $\tau^{[1]}(\bft)=q(\bft)\tau(\bft)$ is the new tau function of the $(M_1,M_2)$-reduction of the 2-BKP hierarchy (\ref{2m1m2bkptaubilinear}).
Notice that $\mathcal{L}=\widetilde{B}_{M_1}^{(1)}+B_{M_2}^{(2)}$,
then by $L_a^{[1]}=T_aL_aT_a^{-1}$ in Theorem \ref{mainqtau}, we can find
\begin{align*}
   \mathcal{L}^{[1]}=\big(T_1[q]\pi_1(\mathcal{L})T_1[q]^{-1}\big)_{1,\geq1}+\big(T_2[q]\pi_2(\mathcal{L})T_2[q]^{-1}\big)_{2,\geq0},
\end{align*}
while $H^{[1]}$ can be obtained by Theorem \ref{mainqtau}. Thus we have finished the proof of {\bf{Proposition \ref{m1m2redction}}}.

If $\widetilde{q}$ is another 2-BKP eigenfunction, then by Corollary \ref{qtauistau}, we can get $\widetilde{q}^{[1]}=q^{-1}\Omega(\widetilde{q},q)
=T_a[q](\widetilde{q})$. So if $\mathcal{L}^{[1]}(\widetilde{q}^{[1]})=\widetilde{c}\widetilde{q}^{[1]}$, 
then by Proposition \ref{m1m2redction}, we can know $\widetilde{q}^{[1]}\tau^{[1]}=\Omega(\widetilde{q},q)\tau$ is the tau function of the $(M_1,M_2)$-reduction of the 2-BKP hierarchy. Here $\mathcal{L}(\widetilde{q})=\widetilde{c}\widetilde{q}$ implies $\mathcal{L}^{[1]}(\widetilde{q}^{[1]})=\widetilde{c}\widetilde{q}^{[1]}$. In fact, according to Proposition \ref{piallam}, there exist $C_{M_2,M_2}, D_{M_1,M_2}\in \E_1$, such that
\begin{align*}
 \mathcal{L}=L_1^{M_1}+C_{M_2,M_2}H,\quad \mathcal{L}^{[1]}=(L_1^{[1]})^{M_1}+D_{M_2,M_2}H.
\end{align*}
Then when $\mathcal{L}(\widetilde{q})=c\widetilde{q}$, we can find by $H(\widetilde{q})=0$ and $\widetilde{q}^{[1]}=T_1[q](\widetilde{q})$ that
\begin{align*}
  \mathcal{L}^{[1]}(\widetilde{q}^{[1]})&=(L_1^{[1]})^{M_1}(\widetilde{q}^{[1]})
  +D_{M_2,M_2}H^{[1]}(\widetilde{q}^{[1]})\\&=T_1L_1^{M_1}(\widetilde{q})
  =\widetilde{c}T_1(\widetilde{q})=\widetilde{c}\widetilde{q}^{[1]}.
\end{align*}
Therefore, we have finished the proof of {\bf{Corollary \ref{mnreduction1}}}.

\section{Additional symmetries and Pfaffian identities of the 2-BKP hierarchy}
In this section, we will consider two applications of the transformations of the 2-BKP tau functions in Theorem \ref{mainqtau}. Firstly, the additional symmetry is one kind of symmetry depending explicitly on the time flows, which can be viewed as the special case of $\Omega(\widetilde{q},q)\tau$ in Corollary \ref{qtauistau}. Then we discuss some Pfaffian identities of the 2-BKP tau functions, where the successive applications of the BKP Darboux operator $T[q]=1-2q^{-1}\partial^{-1}q_{x}$ are used.
\subsection{Additional symmetries of the 2-BKP hierarchy}

Firstly denote
\begin{align*}
    &X_{a}(\lambda)=e^{\xi(t^{(a)},\lambda)}e^{-2\xi(\widetilde{\partial}_{t^{(a)}},\lambda^{-1})}
    ,\quad a=1,2.
 \end{align*}
And define the vertex operators\cite{Date1983,Tu2007,Wu2013}
$X_{ab}(\lambda,\mu)=X_{a}(\mu)X_{b}(-\lambda),$ then we can find
\begin{align*}
  X_{ab}(\lambda,\mu)&=e^{\xi(t^{(a)},\mu)}e^{-2\xi(\widetilde{\partial}_{t^{(a)}},\mu^{-1})}e^{-\xi(t^{(b)},\lambda)}e^{2\xi(\widetilde{\partial}_{t^{(b)}},\lambda^{-1})}
\\&=\varepsilon_{ab}(\lambda,\mu)e^{\xi(t^{(a)},\mu)-\xi(t^{(b)},\lambda)}e^{-2\xi(\widetilde{\partial}_{t^{(a)}},\mu^{-1})+2\xi(\widetilde{\partial}_{t^{(b)}},\lambda^{-1})},
\end{align*}
where $\varepsilon_{ab}(\lambda,\mu)=\frac{\mu+\lambda}{\mu-\lambda}$ if $a=b$, and $\varepsilon_{ab}(\lambda,\mu)=1$ if $a\neq b$.

Let us recall that the 2-BKP wave function $\psi_b(\bft,\lambda)$ is related with the tau function $\tau(\bft)$ by $\psi_b(\mathbf{t},\lambda)=\frac{\tau(\mathbf{t}-2[\lambda^{-1}]_b)}{\tau(\mathbf{t})}e^{\xi(t^{(b)},\lambda)}$. Thus we can find by Theorem \ref{mainqtau} that: $$\tau_b^{[1]}(\bft,\lambda)=X_b(\lambda)\tau(\bft)=\psi_b(\bft,\lambda)\tau(\bft)$$ is still one 2-BKP tau function. Then we have
\begin{align*}
X_{ab}(\lambda,\mu)\tau(\bft)&
=X_{a}(\mu)\tau_b^{[1]}(\bft,-\lambda)=\psi^{[1]}_a(\mathbf{t},\mu)\tau_b^{[1]}(\bft,-\lambda),
\end{align*}
where $\psi_a^{[1]}(\mathbf{t},\mu)=\frac{\tau_b^{[1]}(\mathbf{t}-2[\mu^{-1}]_a,-\lambda)}{\tau_b^{[1]}(\mathbf{t},-\lambda)}e^{\xi(t^{(a)},\mu)}$ can be viewed as the 2-BKP wave function corresponding to $\psi_a(\mathbf{t},\mu)=\frac{\tau(\mathbf{t}-2[\mu^{-1}]_a)}{\tau(\mathbf{t})}e^{\xi(t^{(a)},\mu)}$ under the transformation $\tau(\bft)\rightarrow\tau^{[1]}_b(\bft,-\lambda)=\psi_b(\mathbf{t},-\lambda)\tau(\bft)$. So by Lemma \ref{psi[1]}, we know $\psi_a^{[1]}(\mathbf{t},\mu)=(-1)^a\psi_b(\mathbf{t},-\lambda)^{-1}\Omega\big(\psi_b(\bft,-\lambda),\psi_a(\bft,\mu)\big),$
and 
\begin{align*}
  X_{ab}(\lambda,\mu)\tau(\bft)=(-1)^a\Omega\big(\psi_b(\bft,-\lambda),\psi_a(\bft,\mu)\big)\tau(\bft),
\end{align*}
which is just Lemma \ref{asvmformual}.

Since the SEP $\Omega\big(\psi_a(\bft,\mu),\psi_b(\bft,-\lambda)\big)$ can be up to some constant, thus by Corollary \ref{qtauistau}, we can know
\begin{align*}
\widetilde{\tau}_{ab}=\tau+C_{ab}X_{ab}(\lambda,\mu)\tau
\end{align*}
is also the 2-BKP tau function. So we have proved that the first part of {\bf{Theorem \ref{widetildetau}}}. Further by Proposition \ref{omega12} and Lemma \ref{asvmformual}, we can know
\begin{align*}
\big(e^{-2\xi(\widetilde{\pa}_{t^{(c)}},z^{-1})}-1\big)\frac{X_{ab}(\lambda,\mu)\tau}{\tau}
=&(-1)^{a+c-1}\bigg(\psi_a(\bft,\mu)
  \psi_b(\bft-2[z^{-1}]_c,-\lambda)\\&-\psi_b(\bft,-\lambda)
  \psi_a(\bft-2[z^{-1}]_c,\mu)\bigg).
\end{align*}
If multiply $\psi_c(\bft,z)$ on the both sides and use (\ref{eqigenrel}), we have
\begin{equation}\label{asvmprepaer}
\begin{aligned}
  \psi_c(\bft,z)\big(e^{-2\xi(\widetilde{\pa}_{t^{(c)}},z^{-1})}-1\big)\frac{X_{ab}(\lambda,\mu)\tau}{\tau}
=&(-1)^{a-1}\psi_a(\bft,\mu)\Omega\big(\psi_b(\bft,-\lambda),\psi_c(\bft,z)\big)
  \\&-(-1)^{a-1}\psi_b(\bft,-\lambda)\Omega\big(\psi_a(\bft,\mu),\psi_c(\bft,z)\big).
\end{aligned}
\end{equation}
On the other hand, let us recall that additional flow $\pa_{ab}^*$ on $\psi_c(\bft,z)$ is defined by 
\begin{align*}
 &\partial^*_{ab}\psi_{c}(\mathbf{t},z):=2\Big(\psi_a(\mathbf{t},\mu)\partial_{c}^{-1}\psi_{b,x_c}(\mathbf{t},-\lambda)-\psi_b(\mathbf{t},-\lambda)\partial_{c}^{-1}\psi_{a,x_c}(\mathbf{t},\mu)\Big)\big(\psi_{c}(\mathbf{t},z)\big).
\end{align*}
By $2\pa_c^{-1}(q_{2x_c}q_1)=q_1q_2-(-1)^{c-1}\Omega(q_1,q_2)$ and the direct computation, we can find $\pa_{ab}^*\psi_c(\bft,z)$ coincide with the right side of (\ref{asvmprepaer}). Therefore we finish the proof of {\bf{Theorem \ref{widetildetau}}}.

Next let us discuss the case of the $(M_1,M_2)$-reduction of the 2-BKP hierarchy. 
\begin{lemma}\label{widetildepsi}
  Given 2-BKP tau function $\tau(\bft)$, if we denote 
  $\widetilde{\psi}_c(\bft,z)=\frac{\widetilde{\tau}_{ab}(\bft-2[z^{-1}]_c)}{\widetilde{\tau}_{ab}(\bft)}e^{\xi(t^{(c)},z)}$ for  $\widetilde{\tau}_{ab}=\big(1+C_{ab}X_{ab}(\lambda,\mu)\big)\tau$,
  then
 \begin{align*}
    \widetilde{\psi}_c(\bft,z)&=\psi_c(\bft,z)+\frac{(-1)^{a-1}C_{ab}\Big(
    \psi_a(\bft,\mu)\Omega_{b,c}^{-\lambda,z}(\bft)
    -\psi_b(\bft,-\lambda)\Omega_{a,c}^{\mu,z}(\bft)
    \Big)}
     {1-(-1)^aC_{ab}\Omega_{a,b}^{\mu,-\lambda}(\bft)},
   \end{align*}
    where $\Omega_{a,b}^{\mu,\lambda}(\bft)=\Omega\big(\psi_a(\bft,\mu),\psi_b(\bft,\lambda)\big)$.
\begin{proof}
It can be proved by Lemma \ref{asvmformual} in Subsection \ref{introapplications} and Proposition \ref{omega12} in Subsection \ref{sepdef}.
\end{proof}
\end{lemma}
After the preparation above, let us try to prove Proposition \ref{mnreductionvetex}. Firstly, let us denote $$\mathcal{A}_{abc}^{\mu,\lambda}(\bft,z)=(-1)^{a-1}\frac{C_{ab}\Big(
    \psi_a(\bft,\mu)\Omega_{b,c}^{-\lambda,z}(\bft)
    -\psi_b(\bft,-\lambda)\Omega_{a,c}^{\mu,z}(\bft)
    \Big)}
     {1-(-1)^aC_{ab}\Omega_{a,b}^{\mu,-\lambda}(\bft)},$$
     then $\widetilde{\psi}_c(\bft,z)$ in Lemma \ref{widetildepsi} can be written in to 
      $\widetilde{\psi}_c(\bft,z)=\psi_c(\bft,z)
      +\mathcal{A}_{abc}^{\mu,\lambda}(\bft,z)$.
      
      Next by (\ref{2bkptaubilinear2}), we have
 \begin{equation}\label{rhsofxmn}
     \begin{aligned}
&\sum_{c=1}^{2}{\rm Res}_z(-1)^{c}z^{M_c-1} \widetilde{\psi}_c(\bft,z) \widetilde{\psi}_c(\bft',-z)=\sum_{c=1}^{2}{\rm Res}_z(-1)^{c}z^{M_c-1}\psi_c(\bft,z)\mathcal{A}_{abc}^{\mu,\lambda}(\bft',-z)
     \\&+\sum_{c=1}^{2}{\rm Res}_z(-1)^{c}z^{M_c-1}\psi_c(\bft',-z)\mathcal{A}_{abc}^{\mu,\lambda}(\bft,z)
+\sum_{c=1}^{2}{\rm Res}_z(-1)^{c}z^{M_c-1}\mathcal{A}_{abc}^{\mu,\lambda}(\bft,z)\mathcal{A}_{abc}^{\mu,\lambda}(\bft',-z).
   \end{aligned}
   \end{equation}
Notice that by (\ref{egienspe1}), we have
\begin{align*}
&\sum_{c=1}^{2}{\rm Res}_z(-1)^{c}z^{M_c-1}\psi_c(\bft,z)\mathcal{A}_{abc}^{\mu,\lambda}(\bft',-z)
  \\&\qquad=\frac{2(-1)^{a-1}C_{ab}\Big(
    \psi_a(\bft',\mu)\mathcal{L}\big(\psi_b(\bft,-\lambda)\big)
    -\psi_b(\bft',-\lambda)\mathcal{L}\big(\psi_a(\bft,\mu)\big)
    \Big)}
     {1-(-1)^aC_{ab}\Omega_{a,b}^{\mu,-\lambda}(\bft')},\\
&\sum_{c=1}^{2}{\rm Res}_z(-1)^{c}z^{M_c-1}\psi_c(\bft',-z)\mathcal{A}_{abc}^{\mu,\lambda}(\bft,z)
 \\&\qquad=\frac{2(-1)^{M_c+a-1}C_{ab}\Big(
    \psi_a(\bft,\mu)\mathcal{L}\big(\psi_b(\bft',-\lambda)\big)
    -\psi_b(\bft,-\lambda)\mathcal{L}\big(\psi_a(\bft',\mu)\big)
    \Big)}
     {1-(-1)^aC_{ab}\Omega_{a,b}^{\mu,-\lambda}(\bft)}.
\end{align*}
And by Lemma \ref{Omega12chanes},
\begin{align*}
 \sum_{c=1}^{2}&{\rm Res}_z(-1)^{c}z^{M_c-1}\mathcal{A}_{abc}^{\mu,\lambda}(\bft,z)\mathcal{A}_{abc}^{\mu,\lambda}(\bft',-z)
  =\frac{-2C_{ab}C_{ab}}{\big(1-(-1)^a\Omega_{a,b}^{\mu,-\lambda}(\bft)
  \big)\big(1-(-1)^a\Omega_{a,b}^{\mu,-\lambda}(\bft')
  \big)}\\&\times\bigg(\psi_a(\bft,\mu)\psi_b
  (\bft',-\lambda)\Big(\Omega\big(\psi_b(\bft,-\lambda)
  ,\mathcal{L}\big(\psi_a(\bft,\mu)\big)\big)+(-1)^{M_c}\Omega\big(\psi_a(\bft',\mu)
  ,\mathcal{L}\big(\psi_b(\bft',-\lambda)\big)\big)\Big)
  \\&+\psi_a(\bft',\mu)\psi_b
  (\bft,-\lambda)\Big(\Omega\big(\psi_a(\bft,\mu),\mathcal{L}\big(\psi_b(\bft,-\lambda)
  \big)\big)+(-1)^{M_c}\Omega\big(\psi_b(\bft',-\lambda),\mathcal{L}\big(\psi_a(\bft',\mu)
  \big)\big)\bigg).
\end{align*}
 So finally by Proposition \ref{piallam}, we have 
\begin{align*}
  &{\rm LHS~of~(\ref{rhsofxmn})}=\frac{2(-1)^{a}C_{ab}
  \big(\mu^{M_a}-\lambda^{M_b}
   \big)\Big((-1)^{M_c}\psi_a(\bft',\mu)
   \psi_b(\bft,-\lambda)+
   \psi_a(\bft,\mu)\psi_b(\bft',-\lambda)\Big)}
     {\big(1-(-1)^a\Omega_{a,b}^{\mu,-\lambda}(\bft)\big)
     \big(1-(-1)^a\Omega_{a,b}^{\mu,-\lambda}(\bft')\big)},
\end{align*}
where we notice that $(-1)^{M_a+M_b}=(-1)^{M_c+M_b}=(-1)^{M_c+M_a}=1$. Thus if $\mu^{M_a}=\lambda^{M_b},$  we can find $$\sum_{c=1}^{2}{\rm Res}_z(-1)^{c}z^{M_c-1} \widetilde{\psi}_c(\bft,z) \widetilde{\psi}_c(\bft',-z)=0,$$
which means that $\widetilde{\tau}_{ab}$ is the tau function of the $(M_1,M_2)$-reduction of the 2-BKP hierarchy. Now we have finished the proof of {\bf{Proposition \ref{mnreductionvetex}}}.

\subsection{Pfaffian identities of the 2-BKP tau functions}
In this subsection, we will discuss the successive applications of the transformations in Theorem \ref{mainqtau}. In fact, given $q_1(\bft),\ldots,q_m(\bft)$ to be the 2-BKP eigenfunction, let us define $f^{[m]}$ for the 2-BKP eigenfunction by the following recursion relation
\begin{align*}
  f^{[m]}=q_m^{[m-1]-1}\Omega(f^{[m-1]},q_m^{[m-1]}),\quad m\geq1,
\end{align*}
that is 
\begin{align}\label{fm1}
 f^{[m]}=(-1)^{m+\sum_{i=1}^{m}a_i}T_{a_m}[q_m^{[m-1]}]\cdots T_{a_1}[q_1](f),
\end{align}
where $a_i=1$ or $2$ for $1\leq i\leq m$.

For the 2-BKP tau function $\tau(\bft)$, we can know by Theorem \ref{mainqtau} that
\begin{align}\label{tanm2}
  \tau^{[m]}=q_m^{[m-1]}\cdots q_1\tau
\end{align}
is a new 2-BKP tau function. Then by the result for the BKP Darboux operator in Appendix \ref{app:first} \big(e.g. set $a_1=\ldots=a_m=1$ in (\ref{fm1})\big), we can know
\begin{align}\label{taumofq}
  \frac{\tau^{[m]}}{\tau}=\mathrm{Pf}Q_m(q_m,\ldots,q_1).
\end{align}
Here, $Q_m$ is the $P_m\times P_m$ antisymmetric matrix ($P_m=m$ when $m$ is even, where
$P_m=m+1$ for odd $m$), defined by \begin{align*}
&\Omega_m=\Omega_m(q_m,\ldots, q_1)=\big(\Omega(q_{m+1-i},q_{m+1-j})\big)_{1\leq i,j\leq m},\\
\begin{split}
&Q_m(q_m,\ldots, q_1)=\left \{
\begin{array}{ll}
    \Omega_m                  & m~\rm{is~even},\\
\begin{pmatrix}
\Omega_m & \mathbf{q}\\
    -\mathbf{q}^{T}&0\\
\end{pmatrix},                                 & m~\rm{is~odd}, 
\end{array}
\right.
\end{split}
\end{align*}
and $\mathbf{q}=(q_m,\ldots, q_1)^T$ is a column vector. Here we set $Q_0=1$.
\begin{proposition}
  Notice that $\tau=1$ is a tau function of the 2-BKP hierarchy, the corresponding eigenfunction $q$ should satisfy
  \begin{align}\label{exampletau=1}
   \pa_{a,n}q=\pa_a^n(q),\quad \pa_1\pa_2(q)=0,\quad a=1,2. 
  \end{align}
  Thus we can choose $q$ to be the linear combination of $c_1e^{\xi(t^{(1)},\lambda)}+c_2e^{\xi(t^{(2)},\mu)}$. Therefore $\tau^{[m]}=\mathrm{Pf}Q_m(q_m,\ldots,q_1)$ with $q_i$ satisfying (\ref{exampletau=1}) 
  is the 2-BKP tau function.
\end{proposition}

Next let us assume $m=N_1+N_2$ for $N_1,N_2 \geq0$. When $N_1,N_2 \geq1$, we set
\begin{align*}
  q_i(\bft)=\psi_2(\bft,\mu_{N_2-i+1}),\quad q_j(\bft)=\psi_1(\bft,\lambda_{N_1+N_2 -j+1}),\quad 1\leq i\leq N_2,\quad N_2+1\leq j\leq N_1+N_2.
\end{align*}
When $N_1=0$, let $q_i(\bft)=\psi_2(\bft,\mu_{m-i+1})$ for $1\leq i\leq m$, and when $N_2=0$, let $q_i(\bft)=\psi_1(\bft,\lambda_{m-i+1})$ for $1\leq i\leq m$. Thus we can get by (\ref{taumofq}) that
\begin{align}\label{2bkptaum}
  \frac{\tau^{[m]}}{\tau}=\mathrm{Pf}Q_m\big(\psi_1(\bft,\lambda_{1}),\ldots,
  \psi_1(\bft,\lambda_{N_1}),
  \psi_1(\bft,\mu_{2}),\ldots,\psi_2(\bft,\mu_{N_1})\big).
\end{align}
Let us recall from Remark \ref{remarkpsi2diff} in Subsection \ref{proofth1} that there is a sign difference between $\psi_2^{[1]}(\bft,\mu)$ in Lemma \ref{psi[1]} and $\widetilde{q}^{[1]}$ with $\widetilde{q}=\psi_2(\bft,\mu)$ in Corollary \ref{qtauistau}, under the transformation $\tau\rightarrow q\tau$, that is
\begin{align*}
  \psi_2^{[1]}(\bft,\mu)=-T_1[q]\big(\psi_2(\bft,\mu)\big)=T_2[q]\big(\psi_2(\bft,\mu)\big).
\end{align*}
For general case,
\begin{align*}
  \psi_2^{[m]}(\bft,\mu)=(-1)^{\sum_{i=1}^{m}a_i}T_{a_m}[q_m^{[m-1]}]\cdots T_{a_1}[q_1]\big(\psi_2(\bft,\mu)\big),
\end{align*}
where $a_i=1$ or $2$ for $1\leq i\leq m$.
Therefore by (\ref{tanm2}),
\begin{align*}
  \tau^{[m]}(\bft)=(-1)^{\frac{N_2(N_2-1)}{2}}\psi_1^{[N_1+N_2-1]}
  (\bft,\lambda_{1})\cdots \psi_1^{[N_2]}
  (\bft,\lambda_{N_1})\psi_2^{[N_2-1]}
  (\bft,\mu_{1})\cdots\psi_2
  (\bft,\mu_{N_2})\tau(\bft).
\end{align*}
Further notice that 
$\tau^{[m+1]}(\bft):=\psi_a^{[m]}(\bft,z)\tau^{[m]}(\bft)=X_a(z)\tau^{[m]}(\bft)$.
So we have 
\begin{align*}
 &\tau^{[m]}(\bft)=(-1)^{\frac{N_2(N_2-1)}{2}} X_1(\lambda_{1})\cdots X_1(\lambda_{N_1})X_2(\mu_{1})\cdots X_2(\mu_{N_2})
\tau(\bft)
\\=&\frac{\tau(\mathbf{t}-2\sum_{q=1}^{N_1}[\lambda_q^{-1}]_1-2\sum_{q=1}^{N_2}
    [\mu_q^{-1}]_2)}{\tau(\mathbf{t})} e^{\sum_{q=1}^{N_1}\xi(t^{(1)},\lambda_q)+\sum_{q=1}^{N_2}\xi(t^{(2)},\mu_q)}\prod_{\substack{1\leq i<j\leq N_1\\ 1\leq s<l\leq N_2}}\frac{\lambda_i-\lambda_j}{\lambda_i+\lambda_j}
\frac{\mu_l-\mu_s}{\mu_l+\mu_s}.
\end{align*}

On the other hand, 
\begin{align*}
\begin{split}
&e^{-\sum_{q=1}^{N_1}\xi(t^{(1)},\lambda_q)-\sum_{q=1}^{N_2}\xi(t^{(2)},\mu_q)}
   \mathrm{Pf}Q_m
    \big(\psi_1(\mathbf{t},\lambda_{1}),\ldots,\psi_1(\mathbf{t},\lambda_{N_1})
    ,\psi_2(\mathbf{t},\mu_{1}),\ldots,\psi_2(\mathbf{t},\mu_{N_2})\big)
    \\&\quad=\left \{
\begin{array}{ll}
    \mathrm{Pf}
\begin{pmatrix}
    E & F \\
    -F^{T} & G
\end{pmatrix}_{m\times m},                    &m~\rm{is~even},\\
    \mathrm{Pf}
\begin{pmatrix}
    E & F & C \\
    -F^{T} &G &D \\
   -C^{T} & -D^{T} & 0\\
\end{pmatrix}_{(m+1)\times (m+1)},                                 & m~\rm{is~odd}, 
\end{array}
\right.
\end{split}
\end{align*}
    where 
    \begin{align*}
       &E_{N_1\times N_1}=\bigg(\frac{\lambda_{i}-\lambda_{j}}{\lambda_{i}+\lambda_{j}}\frac{\tau(\mathbf{t}-2[\lambda_{i}^{-1}]_1-2[\lambda_{j}^{-1}]_1)}{\tau(\mathbf{t})}\bigg)
       _{1\leq i,j\leq N_1},\\
       &G_{N_2\times N_2}=\bigg(\frac{\mu_{j}-\mu_{i}}{\mu_{j}+
       \mu_{i}}\frac{\tau(\mathbf{t}-2[\mu_{i}^{-1}]_2
       -2[\mu_{j}^{-1}]_2)}{\tau(\mathbf{t})}\bigg)
       _{1\leq i,j\leq N_2},\\
       &F_{N_1\times N_2}=\bigg(\frac{\tau(\mathbf{t}-2[\lambda_{i}^{-1}]_1-
       2[\mu_{j}^{-1}]_2)}{\tau(\mathbf{t})}\bigg)
       _{1\leq i\leq N_1, 1\leq j\leq N_2},\\     
        &C_{N_1\times 1}=\bigg(\frac{\tau(\mathbf{t}-2[\lambda_{i}^{-1}]_1)}{\tau(\mathbf{t})}\bigg)
       _{1\leq i\leq N_1},
     \quad D_{N_2\times 1 }=\bigg(\frac{\tau(\mathbf{t}-2[\mu_{i}^{-1}]_2)}{\tau(\mathbf{t})}\bigg)_
     {1\leq i\leq N_2}. 
    \end{align*}
where $|\lambda_{N_1}|<|\lambda_{N_1-1}|< \ldots <|\lambda_1|$, $|\mu_{N_2}|<|\mu_{N_2-1}|< \ldots <|\mu_1|$. So finally by (\ref{2bkptaum}) we can prove {\bf{Theorem \ref{pfaffiantau}}}. 
\begin{example}
  When $N_1+N_2=3$ and $N_1+N_2=4$, Theorem \ref{pfaffiantau} will give rise to some 2-BKP addition formulae (or Fay identities). When $N_1=4, N_2=0$, we can obtain the Fay identities of the usual BKP hierarchy in \cite{Tu2007,Zabrodin2021,Zabrodin2025},
  \begin{equation*}
      \begin{aligned}
& \frac{\lambda_1-\lambda_2}{\lambda_1+\lambda_2}
      \frac{\lambda_1-\lambda_3}{\lambda_1+\lambda_3}
      \frac{\lambda_1-\lambda_4}{\lambda_1+\lambda_4}
      \tau(\mathbf{t}-2[\lambda_1^{-1}]_1
      -2[\lambda_2^{-1}]_1-2[\lambda_3^{-1}]_1
      -2[\lambda_4^{-1}]_1)\tau(\mathbf{t})
   \\&=\frac{\lambda_1-\lambda_2}{\lambda_1+\lambda_2}
    \frac{\lambda_2+\lambda_3}{\lambda_2-\lambda_3}
    \frac{\lambda_2+\lambda_4}{\lambda_2-\lambda_4}
   \tau(\mathbf{t}-2[\lambda_1^{-1}]_1
    -2[\lambda_2^{-1}]_1) \tau(\mathbf{t}-2[\lambda_3^{-1}]_1
    -2[\lambda_4^{-1}]_1)
    \\&-\frac{\lambda_1-\lambda_3}{\lambda_1+\lambda_3}
    \frac{\lambda_2+\lambda_3}{\lambda_2-\lambda_3}
    \frac{\lambda_3+\lambda_4}{\lambda_3-\lambda_4}
   \tau(\mathbf{t}-2[\lambda_1^{-1}]_1
    -2[\lambda_3^{-1}]_1) \tau(\mathbf{t}-2[\lambda_2^{-1}]_1
    -2[\lambda_4^{-1}]_1)
    \\&+ \frac{\lambda_1-\lambda_4}{\lambda_1+\lambda_4}
    \frac{\lambda_2+\lambda_4}{\lambda_2-\lambda_4}
    \frac{\lambda_3+\lambda_4}{\lambda_3-\lambda_4}
   \tau(\mathbf{t}-2[\lambda_1^{-1}]_1
    -2[\lambda_4^{-1}]_1) \tau(\mathbf{t}-2[\lambda_2^{-1}]_1
    -2[\lambda_3^{-1}]_1),
      \end{aligned}
    \end{equation*}
    and $N_1=3, N_2=0$ is just the special case of $N_1=4, N_2=0$. For $N_1=3, N_2=1$, we have
    \begin{equation*}
      \begin{aligned}
      & \tau(\mathbf{t}-2[\lambda_1^{-1}]_1
      -2[\lambda_2^{-1}]_1-2[\lambda_3^{-1}]_1-2[\mu_1^{-1}]_2)\tau(\mathbf{t}) 
   \\&=\frac{\lambda_1+\lambda_3}{\lambda_1-\lambda_3}
      \frac{\lambda_2+\lambda_3}{\lambda_2-\lambda_3}
     \tau(\mathbf{t}-2[\lambda_1^{-1}]_1
    -2[\lambda_2^{-1}]_1) \tau(\mathbf{t}-2[\lambda_3^{-1}]_1
    -2[\mu_1^{-1}]_2)
    \\&-\frac{\lambda_1+\lambda_2}{\lambda_1-\lambda_2}
      \frac{\lambda_2+\lambda_3}{\lambda_2-\lambda_3}
     \tau(\mathbf{t}-2[\lambda_1^{-1}]_1
    -2[\lambda_3^{-1}]_1) \tau(\mathbf{t}-2[\lambda_2^{-1}]_1
    -2[\mu_1^{-1}]_2)
    \\&+\frac{\lambda_1+\lambda_2}{\lambda_1-\lambda_2}
      \frac{\lambda_1+\lambda_3}{\lambda_1-\lambda_3}
      \tau(\mathbf{t}-2[\lambda_2^{-1}]_1
    -2[\lambda_3^{-1}]_1) \tau(\mathbf{t}-2[\lambda_1^{-1}]_1
    -2[\mu_1^{-1}]_2).
      \end{aligned}
    \end{equation*}
    If $N_1=2, N_2=2$, we can get
    \begin{equation*}
      \begin{aligned}
      &\tau(\mathbf{t}-2[\lambda_1^{-1}]_1
      -2[\lambda_2^{-1}]_1-2[\mu_1^{-1}]_2
      -2[\mu_2^{-1}]_2)
      \tau(\mathbf{t}) 
   \\&=\tau(\mathbf{t}-2[\lambda_1^{-1}]_1
    -2[\lambda_2^{-1}]_1) \tau(\mathbf{t}-2[\mu_1^{-1}]_2
    -2[\mu_2^{-1}]_2) 
    \\&-\frac{\lambda_1+\lambda_2}{\lambda_1-\lambda_2}
      \frac{\mu_2+\mu_1}{\mu_2-\mu_1}\tau(\mathbf{t}-2[\lambda_1^{-1}]_1
    -2[\mu_1^{-1}]_2) \tau(\mathbf{t}-2[\lambda_2^{-1}]_1
    -2[\mu_2^{-1}]_2) 
    \\&+\frac{\lambda_1+\lambda_2}{\lambda_1-\lambda_2}
      \frac{\mu_2+\mu_1}{\mu_2-\mu_1}\tau(\mathbf{t}-2[\lambda_1^{-1}]_1
    -2[\mu_2^{-1}]_2) \tau(\mathbf{t}-2[\lambda_2^{-1}]_1
    -2[\mu_1^{-1}]_2).
      \end{aligned}
    \end{equation*}
When $N_1=2, N_2=1$, it is
    \begin{equation*}
      \begin{aligned}
      & \tau(\mathbf{t}-2[\lambda_1^{-1}]_1
      -2[\lambda_2^{-1}]_1-2[\mu_1^{-1}]_2)
      \tau(\mathbf{t}) 
   \\&=\tau(\mathbf{t}-2[\lambda_1^{-1}]_1
    -2[\lambda_2^{-1}]_1) \tau(\mathbf{t}-2[\mu_1^{-1}]_2) 
\\&-\frac{\lambda_1+\lambda_2}{\lambda_1-\lambda_2}\tau(\mathbf{t}-2[\lambda_1^{-1}]_1
    -2[\mu_1^{-1}]_2) \tau(\mathbf{t}-2[\lambda_2^{-1}]_1) 
    \\&+\frac{\lambda_1+\lambda_2}{\lambda_1-\lambda_2}\tau(\mathbf{t}-2[\lambda_2^{-1}]_1
    -2[\mu_1^{-1}]_2) \tau(\mathbf{t}-2[\lambda_1^{-1}]_1).
      \end{aligned}
    \end{equation*}
    All these cases are just special cases of addition formulae (or Fay identities) of the 2-BKP hierarchy in \cite{Wu2013,Gao2016}.
\end{example}
\appendix
\section{Pfaffian and BKP Darboux transformations}\label{app:first}
Firstly, the result about Pfaffian and BKP Darboux transformations can be found in many references (e.g., \cite{Cheng2014,Nimmo1995}). Here we give one proof, which can be applied to the 2-BKP hierarchy. Firstly, let us introduce the BKP Darboux operator $T[q]=1-2q^{-1}\pa_xq_x$\cite{Cheng2014,He2007,Nimmo1995} and define the squared eigenfunction potential of the BKP hierarchy \cite{Cheng2010,Loris1999},
$$\widetilde{\Omega}(f,g)=\pa_x^{-1}(f_xg-fg_x).$$ 
By direct computation, we have the lemma below.
\begin{lemma}\label{omegaq1}The following relations hold,
\begin{align*}
  &\widetilde{\Omega}(f_3^{[1]},f_2^{[1]})=\widetilde{\Omega}(f_3,f_2)-f_1^{-1}f_2
  \widetilde{\Omega}(f_3,f_1)+f_1^{-1}f_3
  \widetilde{\Omega}(f_2,f_1),\\
  &\widetilde{\Omega}(f_4^{[2]},f_3^{[2]})\widetilde{\Omega}(f_2,f_1)=\widetilde{\Omega}
  (f_4,f_3)\widetilde{\Omega}(f_2,f_1)-\widetilde{\Omega}(f_4,f_2)
  \widetilde{\Omega}(f_3,f_1)+\widetilde{\Omega}(f_4,f_1)\widetilde{\Omega}(f_3,f_2),
\end{align*}
where $f_i^{[1]}=T[f_1](f_i)=f_1^{-1}\widetilde{\Omega}(f_i,f_1)$, $ f_j^{[2]}=T[f_2^{[1]}](f_j^{[1]})=f_2^{[1]-1}\widetilde{\Omega}(f_j^{[1]},f_2^{[1]})$, $i,j=1,2.$
\end{lemma}

Given an antisymmetric matrix $A=(a_{ij})_{1\leq i,j\leq 2k}$, the Pfaffian \cite{Hirota2004} of $A$ is defined by 
\begin{align}\label{pfdef}
\mathrm{Pf}A=\frac{1}{2^k k!}\sum_{\sigma\in S_{2n}}\mathrm{sgn}(\sigma)\prod_{i=1}^{k}A_{\sigma(2i-1),\sigma(2i)}.
\end{align}
Sometimes, we denote $\mathrm{Pf}A=(1,2,\ldots,2k)$ for convenience. There is one important relations for the Pfaffian\cite{Hirota2004},
\begin{equation}\label{pfaffianrelation}
  \begin{aligned}
 &(a_1,\ldots,a_{2l},1,2,\ldots,2k)(1,2,\ldots,2k)\\&=\sum_{j=2}^{2l}(-1)^j (a_1,a_j,1,2,\ldots,2k)(a_2,\ldots,\widehat{a}_j,\ldots,a_{2l},1,2,\ldots,2k).
\end{aligned}
\end{equation}

Then, given a set of functions $\{f_1,\ldots, f_m\}$ depending on $x$, define an $m\times m$ antisymmetric matrix $\widetilde{\Omega}_m=\widetilde{\Omega}_m(f_m,\ldots, f_1)$ as follows,
\[
(\widetilde{\Omega}_m)_{ij}=\widetilde{\Omega}(f_{m+1-i},f_{m+1-j}),\quad 1\leq i,j\leq m.
\]
Further, we can define a $P_m\times P_m$ matrix $Q_m$ as follows, where $P_m=m$ when $m$ is even, while $P_m=m+1$ for odd $m$. 
\begin{align*}
\begin{split}
&Q_m(f_m,\ldots, f_1)=\left \{
\begin{array}{ll}
    \widetilde{\Omega}_m                   & m\rm{~is~even},\\
\begin{pmatrix}
   \widetilde{\Omega}_m &\mathbf{f}\\
    -\mathbf{f}^{T}& 0\\
\end{pmatrix},                                 & m\rm{~is~odd}, 
\end{array}
\right.
\end{split}
\end{align*}
where $\mathbf{f}=(f_{m},\ldots,f_1)^T$ is a column vector. Here we assume $Q_0=1$.
\begin{lemma}\label{f2npaffian}
If denote $q_i^{[m]}=T[f_m^{[m-1]}]\cdots T[f_2^{[1]}]T[f_1](q_i)$, then
\begin{align}\label{omegafg2n}
\widetilde{\Omega}(q_1^{[2m]},q_2^{[2m]})
  =\frac{\mathrm{Pf}Q_{2m+2}(q_1,q_2,f_{2m},\ldots,f_1)}{\mathrm{Pf}Q_{2m}(f_{2m},\ldots,f_1)}.
\end{align}
\begin{proof}
Firstly, (\ref{omegafg2n}) is obviously correct for $m=0$. Next let us assume (\ref{omegafg2n}) holds for $m=k$. Then by Lemma \ref{omegaq1}, we have
\begin{equation*}
  \begin{aligned}
\widetilde{\Omega}(q_1^{[2k+2]},q_2^{[2k+2]})\widetilde{\Omega}(f_{2k+2}^{[2k]}
,f_{2k+1}^{[2k]})
  =&\widetilde{\Omega}(q_1^{[2k]},q_2^{[2k]})\widetilde{\Omega}
  (f_{2k+2}^{[2k]},f_{2k+1}^{[2k]})-\widetilde{\Omega}(q_1^{[2k]},f_{2k+2}^{[2k]})
\widetilde{\Omega}(q_2^{[2k]},f_{2k+1}^{[2k]})\\&+\widetilde{\Omega}(q_1^{[2k]},f_{2k+1}
^{[2k]})\widetilde{\Omega}(q_2^{[2k]},f_{2k+2}^{[2k]}),
\end{aligned}
\end{equation*}
implying 
\begin{equation*}
  \begin{aligned}
 &\widetilde{\Omega}(q_1^{[2k+2]},q_2^{[2k+2]})\mathrm{Pf}Q_{2k+2}(f_{2k+2},f_{2k+1},f_{2k},
 \ldots,f_1)\mathrm{Pf}Q_{2k}(f_{2k},\ldots,f_1)
  \\&=
  \mathrm{Pf}Q_{2k+2}(q_1,q_2,f_{2k},
 \ldots,f_1)\mathrm{Pf}Q_{2k+2}(f_{2k+2},f_{2k+1},f_{2k},
 \ldots,f_1)
 \\&\quad-\mathrm{Pf}Q_{2k+2}(q_1,f_{2k+2},f_{2k},
 \ldots,f_1)\mathrm{Pf}Q_{2k+2}(q_2,f_{2k+1},f_{2k},
 \ldots,f_1)
 \\&\quad+\mathrm{Pf}Q_{2k+2}(q_1,f_{2k+1},f_{2k},
 \ldots,f_1)\mathrm{Pf}Q_{2k+2}(q_2,f_{2k+2},f_{2k},
 \ldots,f_1),
\end{aligned}
\end{equation*}
Finally (\ref{omegafg2n}) can be proved by (\ref{pfaffianrelation}) for $l=2$.
\end{proof}
 \end{lemma}

\begin{lemma}\label{f2n}The following relation holds
\begin{align*}
  q^{[m]}=\frac{\mathrm{Pf}Q_{m+1}(q,f_{m},\ldots,f_1)}{\mathrm{Pf}Q_{m}(f_{m},\ldots,f_1)},
\end{align*}
and thus $$f_m^{[m-1]}\cdots f_1=\mathrm{Pf}Q_{m}(f_{m},\ldots,f_1).$$
 \begin{proof}
 Firstly, let us prove 
 \begin{align}\label{q2k1}
  q^{[2k]}=\frac{\mathrm{Pf}Q_{2k+1}(q,f_{2k},\ldots,f_1)}{\mathrm{Pf}Q_{2k}(f_{2k},\ldots,f_1)}.
 \end{align}
 Notice that (\ref{q2k1}) is obviously correct for $k=0$. Next let us assume this lemma holds for $\leq k$. We know that 
 $$q^{[2k+2]}=T[f_{2k+2}^{[2k+1]}](q^{2k+1})=f_{2k+2}^{[2k+1]-1}\widetilde{\Omega}(q^{[2k+1]},
 f_{2k+2}^{[2k+1]}).$$
 Further by $f_{2k+2}^{[2k+1]}=T[f_{2k+1}^{[2k]}](f_{2k+2}^{[2k]})$ and the first relation in Lemma \ref{omegaq1},
 \begin{align*}
  q^{[2k+2]}=\frac{f_{2k+1}^{[2k]}
\widetilde{\Omega}(q^{[2k]},f_{2k+2}^{[2k]})
-f_{2k+2}^{[2k]}\widetilde{\Omega}(q^{[2k]},f_{2k+1}^{[2k]})+q^{[2k]} \widetilde{\Omega}(f_{2k+2}^{[2k]},f_{2k+1}^{[2k]})}
{\widetilde{\Omega}(f_{2k+2}^{[2k]},f_{2k+1}^{[2k]})}.
 \end{align*}
 So according to Lemma \ref{f2npaffian} and the assumption for $\leq k$,
 $$q^{[2k+2]}=\frac{A_k}{\mathrm{Pf}Q_{2k+2}(f_{2k+2},f_{2k+1},f_{2k},
 \ldots,f_1)\mathrm{Pf}Q_{2k}(f_{2k},
 \ldots,f_1)},$$
 where 
 \begin{align*}
   A_k=&\mathrm{Pf}Q_{2k+1}(f_{2k+1},
 \ldots,f_1)\mathrm{Pf}Q_{2k+2}(q,f_{2k+2},f_{2k},
 \ldots,f_1)\\&-\mathrm{Pf}Q_{2k+1}(f_{2k+2},f_{2k},
 \ldots,f_1)\mathrm{Pf}Q_{2k+2}(q,f_{2k+1},
 \ldots,f_1)\\&+\mathrm{Pf}Q_{2k+1}(q,f_{2k},
 \ldots,f_1)\mathrm{Pf}Q_{2k+2}(f_{2k+2},
 \ldots,f_1).
 \end{align*}

 If denote $\mathrm{Pf}Q_{2k}(f_{2k},
 \ldots,f_1)=(1,\ldots,2k)$ and set $\mathrm{Pf}Q_{2k+2}(a_1,a_2,f_{2k},
 \ldots,f_1)=(a_1,a_2,1,\ldots,2k)$, $\mathrm{Pf}Q_{2k+1}(a,f_{2k},
 \ldots,f_1)=(a,1,\ldots,2k+1)$. Thus we can find
 \begin{align*}
   A_k=&(f_{2k+1},1,2
 \ldots,2k,2k+1)(q,f_{2k+2},1,2
 \ldots,2k)\\&-(f_{2k+2},1,2
 \ldots,2k,2k+1)(q,f_{2k+1},1,2
 \ldots,2k)\\&+(q,1,2
 \ldots,2k,2k+1)(f_{2k+2},f_{2k+1},1,2
 \ldots,2k).
 \end{align*}
 So by (\ref{pfaffianrelation}) for $l=2$, we can finally obtain
 \begin{align*}
  A_k&=(q,f_{2k+2},f_{2k+1},1,2
 \ldots,2k,2k+1)(1,2
 \ldots,2k)\\&=\mathrm{Pf}Q_{2k+3}(q,f_{2k+2},
 \ldots,f_1)\mathrm{Pf}Q_{2k}(f_{2k},
 \ldots,f_1),
 \end{align*}
 which means 
   \begin{align*}
  q^{[2k+2]}=\frac{\mathrm{Pf}Q_{2k+3}(q,f_{2k+2},
 \ldots,f_1)}{\mathrm{Pf}Q_{2k+2}(f_{2k+2},
 \ldots,f_1)}.
\end{align*}
Thus we have proved (\ref{q2k1}).

On the other hand, by (\ref{omegafg2n}) and (\ref{q2k1}), we can find 
\begin{align}\label{q2k+1}
  q^{[2k+1]}=T[f_{2k+1}^{2k}](q^{2k})=\frac{\mathrm{Pf}Q_{2k+2}(q_1,f_{2k+1},\ldots,f_1)}
  {\mathrm{Pf}Q_{2k+1}(f_{2k+1}
  ,\ldots,f_1)}.
\end{align}
From (\ref{q2k1}) and (\ref{q2k+1}), we can know this lemma is correct.
 \end{proof}
\end{lemma}

\begin{corollary}\label{fnexpress}
  If we define $T_i=T_i[\frac{\mathrm{Pf}Q_{i}}{\mathrm{Pf}Q_{i-1}}]$ with $Q_i=Q_i(f_i\ldots,f_1)$, then
  \begin{align*}
T_{m}\cdots T_1(q_1)=\frac{\mathrm{Pf}Q_{m+1}(q,f_{m},\ldots,f_{1})}{\mathrm{Pf}Q_{m}(f_{m},\ldots,f_{1})}. \end{align*}
\end{corollary}

\section*{Acknowledgement}
This article is dedicated to Professor Ke Wu in Capital Normal University in celebration of his 80th birthday. And this work is supported by National Natural Science Foundation of China (Grant Nos. 12571271, 12171472, 12261072).\\

\section*{Conflict of Interest}
The authors have no conflicts to disclose.\\

\section*{Data availability}
     Data sharing is not applicable to this article as no new data were created or analyzed in this study.

\end{document}